\documentclass[11pt]{article}

\usepackage{amssymb}
\usepackage{amsmath}
\usepackage{amsthm}
\usepackage[dvips]{graphics,color}
\usepackage{epsfig}

\usepackage{a4wide}

\def\E{\mathbb{E}}
\def\var{\text{Var}}

\begin{document}

\title{Dynamic generalized linear models for non-Gaussian time series
forecasting}

\author{K. Triantafyllopoulos\footnote{Department of
Probability and Statistics, Hicks Building, University of Sheffield,
Sheffield S3 7RH, UK, email: {\tt
k.triantafyllopoulos@sheffield.ac.uk}}}

\date{\today}

\maketitle

\begin{abstract}

The purpose of this paper is to provide a discussion, with
illustrating examples, on Bayesian forecasting for dynamic
generalized linear models (DGLMs). Adopting approximate Bayesian
analysis, based on conjugate forms and on Bayes linear estimation,
we describe the theoretical framework and then we provide detailed
examples of response distributions, including binomial, Poisson,
negative binomial, geometric, normal, log-normal, gamma,
exponential, Weibull, Pareto, beta, and inverse Gaussian. We give
numerical illustrations for all distributions (except for the
normal). Putting together all the above distributions, we give a
unified Bayesian approach to non-Gaussian time series analysis,
with applications from finance and medicine to biology and the
behavioural sciences. Throughout the models we discuss Bayesian
forecasting and, for each model, we derive the multi-step forecast
mean. Finally, we describe model assessment using the likelihood
function, and Bayesian model monitoring.

\textit{Some key words:} Bayesian forecasting, non-Gaussian time
series, dynamic generalized linear model, state space, Kalman
filter.

\end{abstract}

\section{Introduction}\label{introduction}

In the past three decades non-Gaussian time series have attracted a
lot of interest, see e.g. Cox (1981), Kaufmann (1987), Kitagawa
(1987), Shephard and Pitt (1997), and Durbin and Koopman (2000),
among others. In the context of regression modelling, generalized
linear models (McCullagh and Nelder, 1989; Dobson, 2002) offer a
solid theoretical basis for statistical analysis of independent
non-normal data. A general framework for dealing with time series
data is the dynamic generalized linear model (DGLM), which considers
generalized linear modelling with time-varying parameters and hence
it is capable to model time series data for a wide range of response
distributions. DGLMs have been widely adopted for non-normal time
series data, see e.g. West {\it et al.} (1985), Gamerman and West
(1987), Fahrmeir (1987), Fr\"{u}hwirth-Schnatter, S. (1994), Lindsey
and Lambert (1995), Chiogna and Gaetan (2002), Hemming and Shaw
(2002), Godolphin and Triantafyllopoulos (2006), and Gamerman (1991,
1998). Dynamic generalized linear models are reported in detail in
the monographs of West and Harrison (1997, Chapter 14), Fahrmeir and
Tutz (2001, Chapter 8), and Kedem and Fokianos (2002, Chapter 6).

In this paper we propose a unified treatment of DGLMs that includes
approximate Bayesian inference and multi-step forecasting. In this
to end we adopt the estimation approach of West {\it et al.} (1985),
but we extend it as far as model diagnostics and forecasting are
concerned. In particular, we discuss likelihood-based model
assessment as well as Bayesian model monitoring. In the literature,
discussion on the DGLMs is usually restricted to the binomial and
the Poisson models, see e.g. Fahrmeir and Tutz (2001, Chapter 8).
Even for these response distributions, discussion is limited on
estimation, while forecasting and in particular multi-step
forecasting does not appear to have received much attention. We
provide detailed examples of many distributions, including binomial,
Poisson, negative binomial, geometric, normal, log-normal, gamma,
exponential, Weibull, Pareto, two special cases of the beta, and
inverse Gaussian. We give numerical illustrations for all
distributions, except for the normal (for which one can find
numerous illustrations in the time series literature) using real and
simulated data.

The paper is organized as follows. In Section \ref{dglm} we discuss
Bayesian inference of DGLMs. Section \ref{examples} commences by
considering several examples, where the response time series follows
a particular distribution. Section \ref{discussion} gives concluding
comments. The appendix includes some proofs of arguments in Section
\ref{examples}.

\section{Dynamic generalized linear models}\label{dglm}

\subsection{Model definition}

Suppose that the time series $\{y_t\}$ is generated from a
probability distribution, which is a member of the exponential
family of distributions, that is
\begin{equation}\label{exp}
p(y_t|\gamma_t) = \exp\left( \frac{1}{a(\phi_t)} \left(
z(y_t)\gamma_t - b(\gamma_t)\right) \right) c(y_t,\phi_t),
\end{equation}
where $\gamma_t$, known as the natural parameter, is the parameter
of interest and other parameters that can be linked to $\phi_t$,
$a(.)$, $b(.)$ and $c(.,.)$ are usually referred to as nuisance
parameters or hyperparameters. The functions $a(.)$, $b(.)$ and
$c(.,.)$ are assumed known, $\phi_t,a(\phi_t),c(y_t,\phi_t)>0$,
$b(\gamma_t)$ is twice differentiable and according to Dobson (2002,
\S3.3)
$$
\E(z(y_t)|\gamma_t)=\frac{\,db(\gamma_t)}{\,d\gamma_t} \quad
\textrm{and} \quad \var(z(y_t)|\gamma_t) =
\frac{a(\phi_t)\,d^2b(\gamma_t)}{ \gamma_t^2}.
$$
The function $z(.)$ is usually a simple function in $y_t$ and in
many cases it is the identity function; an exception of this is the
binomial distribution. If $z(y_t)=y_t$, distribution (\ref{exp}) is
said to be in the {\it canonical} or {\it standard} form. Dobson
(2002, \S3.3) gives expressions of the score statistics and the
information matrix, although the consideration of these may not be
necessary for Bayesian inference.

The idea of generalized linear modelling is to use a non-linear
function $g(.)$, which maps $\mu_t=\E(y_t|\gamma_t)$ to the linear
predictor $\eta_t$; this function is known as link function. If
$g(\mu_t)=\gamma_t$, this is referred to as {\it canonical link},
but other links may be more useful in applications (see e.g. the
inverse Gaussian example in Section \ref{continuous}). In GLM
theory, $\eta_t$ is modelled as a linear model, but in DGLM theory,
the linear predictor is replaced by a state space model, i.e.
$$
g(\mu_t)=\eta_t=F_t'\theta_t \quad \textrm{and} \quad \theta_t=
G_t\theta_{t-1}+\omega_t,
$$
where $F_t$ is a $d\times 1$ design vector, $G_t$ is a $d\times d$
evolution matrix, $\theta_t$ is a $d\times 1$ random vector and
$\omega_t$ is an innovation vector, with zero mean and some known
covariance matrix $\Omega_t$. It is assumed that $\omega_t$ is
uncorrelated of $\omega_s$ (for $t\neq s$) and $\omega_t$ is
uncorrelated of $\theta_0$, for all $t$. It is obvious that if one
sets $G_t=I_p$ (the $d\times d$ identity matrix) and $\omega_t=0$
(i.e. its covariance matrix is the zero matrix), then the above
model is reduced to a usual GLM.

For the examples of Section \ref{examples} we consider simple state
space models, which assume that $F_t=F$, $G_t=G$, $\Omega_t=\Omega$
are time-invariant. However, in the next sections, we present
Bayesian inference and forecasting for time-varying $F_t$, $G_t$,
$\Omega_t$ in order to cover the general situation.

\subsection{Bayesian inference}

Suppose that we have data $y_1,\ldots,y_T$ and we form the
information set $y^t=\{y_1,\ldots,y_t\}$, for $t=1,\ldots,T$. At
time $t-1$ we assume that the posterior mean vector and covariance
matrix of $\theta_{t-1}$ are $m_{t-1}$ and $P_{t-1}$, respectively,
and we write $\theta_{t-1}|y^{t-1}\sim (m_{t-1},P_{t-1})$. Then from
$\theta_t=G_t\theta_{t-1}+\omega_t$, it follows that
$\theta_t|y^{t-1}\sim (h_t, R_t)$, where $h_t=G_tm_{t-1}$ and
$R_t=G_tP_{t-1}G_t'+\Omega_t$.

The next step is to form the prior mean and variance of $\eta_t$ and
$\theta_t$, that is
\begin{equation}\label{prior:p2}
\left.\left[ \begin{array}{c} \eta_t \\ \theta_t \end{array}\right]
\right|y^{t-1} \sim  \left(\left[ \begin{array}{c} f_t \\ h_t
\end{array}\right], \left[ \begin{array}{cc} q_t &
F_t'R_t \\ R_tF_t & R_t
\end{array} \right] \right),
\end{equation}
where $f_t=F_t'h_t$ and $q_t=F_t'R_tF_t$. The quantities $f_t$ and
$q_t$ are the forecast mean and variance of $\eta_t$.

In order to proceed with Bayesian inference, we assume the conjugate
prior of $\gamma_t$, so that
\begin{equation}\label{prior:g1}
p(\gamma_t|y^{t-1})=\kappa(r_t,s_t)\exp(r_t\gamma_t-s_tb(\gamma_t)),
\end{equation}
for some known $r_t$ and $s_t$. These parameters can be found from
$g(\mu_t)=\eta_t$ and $f_t=\E(\eta_t|y^{t-1})$,
$q_t=\var(\eta_t|y^{t-1})$, which are known from (\ref{prior:p2}).
The normalizing constant $\kappa(.,.)$ can be found by
$$
\kappa(r_t,s_t)=\left(\int
\exp(r_t\gamma_t-s_tb(\gamma_t))\,d\gamma_t \right)^{-1},
$$
where the integral is Lebesque integral, so that it includes
summation / integration of discrete / continuous variables. We note
that in most of the cases, the above distribution will be
recognizable (e.g. gamma, beta, normal) and so there is no need of
evaluating the above integral. One example that this is not the case
is the inverse Gaussian distribution (see Section \ref{continuous}).

Then observing $y_t$, the posterior distribution of $\gamma_t$ is
\begin{eqnarray}
p(\gamma_t|y^t) & = & \frac{p(y_t|\gamma_t) p(\gamma_t|y^{t-1})} {
\int p(y_t|\gamma_t)p(\gamma_t|y^{t-1})\,d\gamma_t} \nonumber
\\ &=&
\kappa\left(r_t+\frac{z(y_t)}{a(\phi_t)},s_t+\frac{1}{a(\phi_t)}\right)
\exp\left(\left(r_t+\frac{z(y_t)}{a(\phi_t)}\right) \gamma_t -
\left(s_t+\frac{1}{a(\phi_t)}\right)b(\gamma_t)
\right).\label{post:g1}
\end{eqnarray}
In many situations we are interested in parameters that are given as
functions of $\gamma_t$. In such cases we derive the prior/posterior
distributions of $\gamma_t$ as above and then we apply a
transformation to obtain the prior/posterior distribution of the
parameter in interest. The examples of Section \ref{examples} are
illuminative.

Finally, the posterior mean vector and covariance matrix of
$\theta_t$ are approximately given by
\begin{equation}\label{post:th1}
\theta_t|y^t \sim (m_t,P_t),
\end{equation}
with
$$
m_t=h_t+R_tF_t(f_t^*-f_t)/q_t \quad \textrm{and} \quad P_t= R_t -
R_t F_tF_t' R_t (1-q_t^*/q_t)/q_t,
$$
where $f_t^*=\E(\eta_t|y^t)$ and $q_t^*=\E(\eta_t|y^t)$ can be found
from $g(\mu_t)=\eta_t$ and the posterior (\ref{post:g1}). The priors
(\ref{prior:p2}), (\ref{prior:g1}) and the posteriors
(\ref{post:g1}), (\ref{post:th1}) provide an algorithm for
estimation, for any $t=1,\ldots,T$. For a proof of the above
algorithm the reader is referred to West {\it et al.} (1985).

An alternative approach for the specification of $r_t$ and $s_t$ is
to make use of {\it power discounting} and this is briefly discussed
next. The idea of power discounting stem in the work of Smith
(1979); power discounting is a method of obtaining the prior
distribution at time $t+1$, from the posterior distribution at time
$t$. Here we consider a minor extension of the method by replacing
$t+1$ by $t+\ell$, for some positive integer $\ell$. Then, according
to the principle of power discounting, the prior distribution at
time $t+\ell$ is proportional to $(p(\gamma_t|y^t))^\delta$, where
$\delta$ is a discount factor. Thus we write
$$
p(\gamma_{t+\ell}|y^t) \propto (p(\gamma_t)|y^t)^\delta, \quad
\textrm{for} \quad 0<\delta<1.
$$
This ensures that the prior distribution of $\gamma_{t+\ell}$ is
flatter than the posterior distribution of $\gamma_t$. The above
procedure assumes that $r_t(\ell)=r_{t+1}$ and $s_t(\ell)=s_{t+1}$,
which implicitly assumes a random walk type evolution of the
posterior/prior updating, in the sense that Bayes decisions in the
interval $(t,t+\ell)$ remain constant, while the respective expected
loss (under step loss functions) increase (Smith, 1979).

\subsection{Bayesian forecasting and model assessment}

Suppose that the time series $\{y_t\}$ is generated by density
(\ref{exp}) and let $y^t$ be the information set up to time $t$.
Then the $\ell$-step forecast distribution of $y_{t+\ell}$ is
\begin{equation}\label{eq:for}
p(y_{t+\ell}|y^t) = \int p(y_{t+\ell}|\gamma_{t+\ell})
p(\gamma_{t+\ell}|y^t)\,d\gamma_{t+\ell} = \frac{
\kappa(r_t(\ell),s_t(\ell)) c(y_{t+\ell},\phi_{t+\ell}) }{
\kappa\left(r_t(\ell)+\frac{z(y_{t+\ell})}{a(\phi_{t+\ell})},
s_t(\ell)+\frac{1}{a(\phi_{t+\ell})}\right)},
\end{equation}
where $r_t(\ell)$ and $s_t(\ell)$ are evaluated from $f_t(\ell)$ and
$q_t(\ell)$, the mean and variance of $\eta_{t+\ell}|y^t$, and the
distribution of $\gamma_{t+\ell}|y^t$, which takes a similar form as
the distribution of $\gamma_t|y^{t-1}$.

Model assessment can be done via the likelihood function, residual
analysis, and Bayesian model comparison, e.g. based on Bayes
factors. The likelihood function of $\gamma_1,\ldots,\gamma_T$,
based on information $y^T$ is
$$
L(\gamma_1,\ldots,\gamma_T;y^T)=\prod_{t=1}^T
p(y_t|\gamma_t)p(\gamma_t|\gamma_{t-1}),
$$
where the first probability in the product is the distribution
(\ref{exp}) and the second indicates the evolution of $\gamma_t$,
given $\gamma_{t-1}$. Then the log-likelihood function is
\begin{equation}\label{logl}
\ell(\gamma_1,\ldots,\gamma_T;y^T)= \sum_{t=1}^T \left(
\frac{1}{a(\phi_t)} (z(y_t)\gamma_t-b(\gamma_t)) + \log
c(y_t,\phi_t) \right) + \sum_{t=1}^T \log p(\gamma_t|\gamma_{t-1}).
\end{equation}
The likelihood function can be used as a means of model comparison
(for example looking at two model specifications, which differ in
some quantitative parts, we choose the model that has larger
likelihood). For model assessment the likelihood function can be
used in order to choose some hyperparameters (discount factors, or
nuisance parameters) so that the likelihood function is maximized in
terms of these hyperparameters. The evaluation of (\ref{logl})
requires the distribution $p(\gamma_t|\gamma_{t-1})$. This depends
on the state space model for $\eta_t$ used. In the examples of
Section \ref{examples} we look at these probabilities, based mainly
on Gaussian random walk evolutions for $\eta_t$, but also we
consider a linear trend model for $\eta_t$. Note that the
consideration of $\omega_t$ following a Gaussian distribution does
not imply that $\theta_t|y^t$ follows a Gaussian distribution too,
since the distribution of $\theta_0$ may not be Gaussian.

For the sequential calculation of the Bayes factors (which for
Gaussian responses are discussed in Salvador and Gargallo, 2005), a
typical setting suggests the formation of two models $\mathcal{M}_1$
and $\mathcal{M}_2$, which differ in some quantitative aspects, e.g.
some hyperparameters. Then, the cumulative Bayes factor of
$\mathcal{M}_1$ against $\mathcal{M}_2$ is defined by
\begin{equation}\label{bf1}
H_t(k) =  \frac{ p(y_{t}, \ldots, y_{t-k+1} | y^{t-k}, \mathcal{M}_1
) }{ p(y_{t}, \ldots, y_{t-k+1} | y^{t-k}, \mathcal{M}_2 ) } =
H_{t-1}(k-1) H_t(1) = \prod_{i=t-k+1}^t H_i(1)
\end{equation}
where $H_1(1)=H_t(0)=1$, for all $t$, and $p(y_{t}, \ldots,
y_{t-k+1} | y^{t-k}, \mathcal{M}_j )$ denotes the joint distribution
of $y_{t},\ldots,y_{t-k+1}$, given $y^{t-k}$, for some integer
$0<k<t$ and $j=1,2$. Then preference of model 1 would imply larger
forecast distribution of this model (or $H_t(k)>1$); likewise
preference of model 2 would imply $H_t(k)<1$; $H_t(k)=1$ implies
that the two models are probabilistically equivalent in the sense
they provide the same forecast distributions.

\section{Examples}\label{examples}

\subsection{Discrete distributions for the response $y_t$}\label{discrete}

\subsubsection{Binomial}

The binomial distribution (Johnson {\it et al.}, 2005) is perhaps
the most popular discrete distribution. It is typically generated as
the sum of independent success/failure bernoulli trials and in the
context of generalized linear modelling is associated with logistic
regression (Dobson, 2002).

Consider a discrete-valued time series $\{y_t\}$, which, for a given
probability $\pi_t$, follows the binomial distribution
$$
p(y_t|\pi_t)=\binom {n_t}{y_t} \pi_t ^{y_t} (1-\pi_t)^{n_t-y_t},
\quad y_t=0,1,2,\ldots,n_t; \quad n_t=1,2,\ldots; \quad 0<\pi_t<1,
$$
where $\binom {n_t}{y_t}$ denotes the binomial coefficient. It is
easy to verify that the above distribution is of the form
(\ref{exp}) with $z(y_t)=y_t/n_t$, $\gamma_t=\log \pi_t/(1-\pi_t)$,
$a(\phi_t)=\phi_t^{-1}=n_t^{-1}$, $b(\gamma_t)=\log
(1+\textrm{exp}(\gamma_t))$, and $c(\gamma_t,\phi_t)=\binom
{n_t}{y_t}$. The logarithmic, known also as logit, link
$\eta_t=g(\mu_t)=\gamma_t=\log \pi_t/(1-\pi_t)$ maps $\pi_t$ to the
linear predictor $\eta_t$, which with the setting
$\eta_t=F'\theta_t$ and $\theta_t=G\theta_{t-1}+\omega_t$, generates
the dynamic evolution of the model.

The prior of $\pi_t|y^{t-1}$, follows by the prior of
$\gamma_t|y^{t-1}$ and the transformation $\gamma_t=\log
\pi_t/(1-\pi_t)$ as beta distribution $\pi_t|y^{t-1}\sim
B(r_t,s_t-r_t)$, with density
$$
p(\pi_t|y^{t-1})=\frac{\Gamma(s_t)}{\Gamma(r_t)\Gamma(s_t-r_t)}
\pi_t^{r_t-1} (1-\pi_t)^{s_t-r_t-1},
$$
where $\Gamma(.)$ denotes the gamma function and $s_t>r_t>0$. Then,
observing $y_t$, the posterior of $\pi_t|y^t$ is $\pi_t|y^t\sim
B(r_t+y_t,s_t+n_t-r_t-y_t)$.

In the appendix it is shown that, with $f_t$ and $q_t$ the prior
mean and variance of $\eta_t$, an approximation of $r_t$ and $s_t$
is given by
\begin{equation}\label{eq:binom:rt}
r_t=\frac{1+\textrm{exp}(f_t)}{q_t} \quad \textrm{and} \quad
s_t=\frac{2+\textrm{exp}(f_t)+\textrm{exp}(-f_t)}{q_t}.
\end{equation}
In order to proceed with the posterior moments of $\theta_t|y^t$ as
in (\ref{post:th1}), we can see that
$$
f_t^*=\psi(r_t+y_t)-\psi(s_t-r_t+n_t-y_t) \quad \textrm{and} \quad
q_t^*= \left.\frac{\,d\psi(x)}{\,dx}\right|_{x=r_t+y_t}+
\left.\frac{\,d\psi(x)}{\,dx}\right|_{x=s_t+n_t-y_t},
$$
where $\psi(.)$ denotes the digamma function (see the Poisson
example and the appendix). In the appendix approximations of
$\psi(.)$ and of its first derivative (also known as trigamma
function) are given. These definitions as well as the parameters of
the beta prior are slightly different from the ones obtained by West
and Harrison (1997), as these authors use a different
parameterization, which does not appear to be consistent with the
prior/posterior updating.

Given information $y^t$, the $\ell$-step forecast distribution is
obtained by first noting that
\begin{equation}\label{eq:binom2}
\pi_{t+\ell}|y^t\sim B(r_t(\ell),s_t(\ell)-r_t(\ell)),
\end{equation}
where $r_t(\ell)$ and $s_t(\ell)$ are given by $r_t$ and $s_t$, if
$f_t$ and $q_t$ are replaced by $f_t(\ell)=\E(\eta_{t+\ell}|y^t)$
and $q_t(\ell)=\var(\eta_{t+\ell}|y^t)$, which are calculated
routinely by the Kalman filter (see Section \ref{dglm}). Then the
$\ell$-step forecast distribution is given by
\begin{eqnarray*}
p(y_{t+\ell}|y^t) &=& \frac{ \Gamma(s_t(\ell)) } { \Gamma(r_t(\ell))
\Gamma(s_t(\ell)-r_t(\ell))\Gamma(s_t(\ell)+n_{t+\ell}) } \\ &&
\times \frac{1}{n_{t+\ell}} \binom {n_{t+\ell}} {y_{t+\ell}}
\Gamma(r_t(\ell)+y_{t+\ell})
\Gamma(s_t(\ell)-r_t(\ell)+n_{t+\ell}-y_{t+\ell} ).
\end{eqnarray*}
We can use conditional expectations in order to calculate the
forecast mean and variance, i.e.
$$
y_t(\ell)=\E(y_{t+\ell}|y^t)=\E(\E(y_{t+\ell}|\pi_{t+\ell})|y^t) =
\frac{ n_{t+\ell} (r_t(\ell)+1) }{ r_t(\ell) + s_t(\ell) + 1}
$$
and
\begin{eqnarray*}
\var(y_{t+\ell}|y^t) &=& \E(\var(y_{t+\ell}|\pi_{t+\ell})|y^t) +
\var(\E(y_{t+\ell}|\pi_{t+\ell})) \\ &=& \frac{ n_{t+\ell}
(r_t(\ell)+1) }{ r_t(\ell) + s_t(\ell) + 1} - \frac{ n_{t+\ell}
(r_t(\ell) +1) (r_t(\ell)+2) }{ (r_t(\ell)+s_t(\ell) +1) (r_t(\ell)
+ s_t(\ell) + 2) } \\ && + \frac{ n_{t+\ell}^2 (r_t(\ell)+1)
s_t(\ell) }{ (r_t(\ell)+s_t(\ell) +1)^2 (r_t(\ell) + s_t(\ell) + 2)
}
\end{eqnarray*}

For the specification of $r_t$ and $s_t$, we can alternatively use
power discounting (see Section \ref{dglm}). This yields
$$
r_{t+1}=\delta r_t + \delta y_t +1-\delta \quad \textrm{and} \quad
s_{t+1} = \delta s_t+\delta n_t + 2-\delta,
$$
where $\delta$ is a discount factor and $r_0,s_0$ are initially
given.

For the evolution of $\eta_t$ via $\theta_t$, the obvious setting is
the random walk, which sets $\eta_t=\theta_t=\theta_{t-1}+\omega_t$.
From the logit link we have $\pi_t/(1-\pi_t)=\textrm{exp}(\theta_t)$
and so the evolution of $\theta_t$ yields
$$
\pi_t=\frac{ \textrm{exp}(\omega_t)\pi_{t-1}} {
1-\pi_{t-1}+\textrm{exp}(\omega_t)\pi_{t-1}},
$$
which gives the evolution of $\pi_t$, given $\pi_{t-1}$, as a
function of the Gaussian shock $\omega_t$. Then the distribution of
$\pi_t|\pi_{t-1}$ is
$$
p(\pi_t|\pi_{t-1})=\frac{1}{\sqrt{2\pi\Omega} \pi_t(1-\pi_t)}
\exp\left(-\frac{1}{2\Omega}\left(\log\frac{\pi_t(1-\pi_{t-1})}{
\pi_{t-1}(1-\pi_t)}\right)^2\right)
$$
and so from (\ref{logl}) the log-likelihood function is
\begin{eqnarray*}
\ell(\pi_1,\ldots,\pi_T;y^T) &=& \sum_{t=1}^T \bigg( y_t \log\pi_t
-y_t\log(1-\pi_t)+n_t\log(1-\pi_t)+\log\binom {n_t}{y_t} \\ &&
-\log\sqrt{2\pi\Omega}\pi_t(1-\pi_t)
-\frac{1}{2\Omega}\left(\log\frac{\pi_t(1-\pi_{t-1})}{\pi_{t-1}(1-\pi_t)}\right)^2
\bigg)
\end{eqnarray*}
The Bayes factors are easily computed from (\ref{bf1}) and the
forecast distribution $p(y_{t+\ell}|y^t)$.

If we use a linear trend evolution on $\theta_t$, we can specify
$$
\eta_t=[1~0]\left[\begin{array}{c} \theta_{1t} \\
\theta_{2t}\end{array}\right] \quad \textrm{and} \quad \left[\begin{array}{c} \theta_{1t} \\
\theta_{2t}\end{array}\right] = \left[\begin{array}{cc} 1 & 1 \\ 0 &
1
\end{array}\right] \left[\begin{array}{c} \theta_{1,t-1} \\
\theta_{2,t-1}\end{array}\right] + \left[\begin{array}{c} \omega_{1t} \\
\omega_{2t}\end{array}\right].
$$
Here $\theta_t=[\theta_{1t}~\theta_{2t}]'$ is a 2-dimensional random
vector and $\omega_t=[\omega_{1t}~\omega_{2t}]'$ follows a bivariate
normal distribution with zero mean vector and some known covariance
matrix. Then, conditional on $\pi_{t-1}$, from the logit link
function we can recover the relationship of $\pi_t$ as
$$
\pi_t=\frac{ \textrm{exp}(\theta_{2,0} + \sum_{i=1}^t
\omega_{2i}+\omega_{1t} ) \pi_{t-1} } { 1- \pi_{t-1} + \textrm{exp}
(\theta_{2,0}+\sum_{i=1}^t\omega_{2i} +\omega_{1t})\pi_{t-1}}.
$$

\begin{figure}[t]
 \epsfig{file=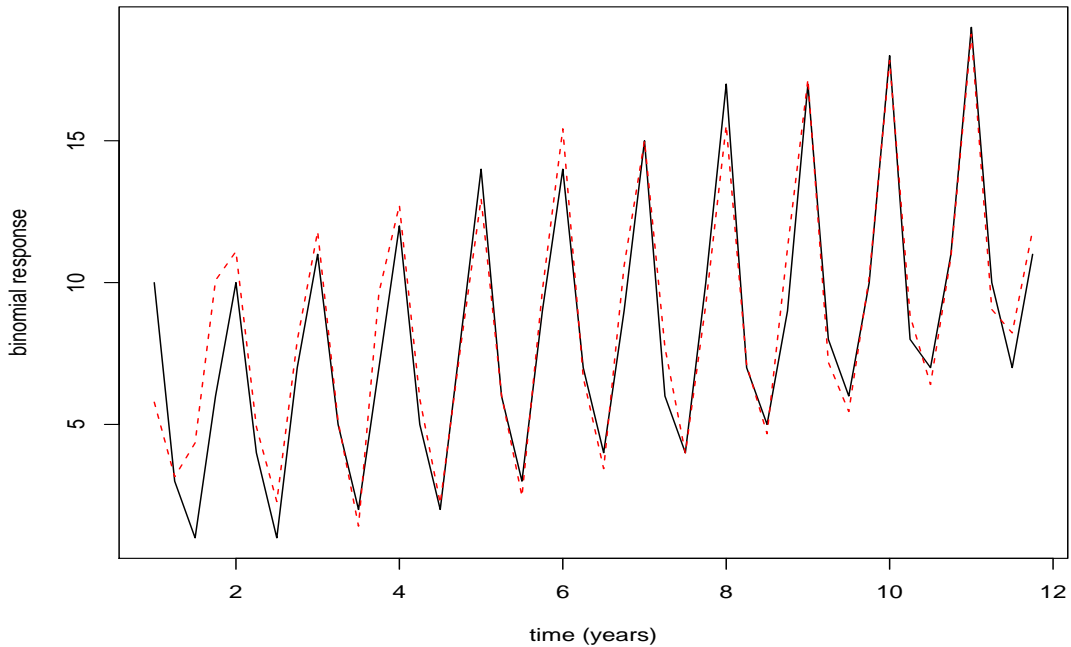, height=10cm, width=15cm}
 \caption{Binomial data of 25 Bernoulli trials (solid line) and
 one-step forecast mean (dashed line).}\label{fig1}
\end{figure}

To illustrate the binomial model, we consider the data of Godolphin
and Triantafyllopoulos (2006), consisting of quarterly binomial data
over a period of 11 years. In each quarter $n_t=25$ Bernoulli trials
are performed and $y_t$, the number of successes, is recorded. The
data, which are plotted in Figure \ref{fig1}, show a clear
seasonality and therefore, modelling this data with GLMs is
inappropriate. The data exhibit a trend/periodic pattern, which can
be modelled with a DGLM, by setting $\eta_t=F'\theta_t$ and
$\theta_t=G\theta_{t-1}+\omega_t$, where the design vector $F$ has
dimension $5\times 1$ and the $5\times 5$ evolution matrix $G$
comprises a linear trend component and a seasonal component. One way
to do this is by applying the trend / full harmonic state space
model
$$
F=\left[ \begin{array}{c} 1 \\ 0 \\ 1 \\ 0 \\ 1
\end{array}\right] \quad \textrm{and} \quad G=\left[ \begin{array}{ccccc} 1 & 1 & 0 &
0 & 0 \\ 0 & 1 & 0 & 0 & 0 \\ 0 & 0 & \cos (\pi/2) & \sin (\pi/2) & 0 \\
0 & 0 & -\sin (\pi/2) & \cos (\pi/2) & 0 \\ 0 & 0 & 0 & 0 & -1
\end{array}\right],
$$
where $G$ is a block diagonal matrix, comprising the linear trend
component and the seasonal component, for the latter of which, with
a cycle of $c=4$, we have $h=c/2=2$ harmonics and the frequencies
are $\omega=2\pi/4=\pi/2$ for harmonic 1 and $\omega=4\pi/4=\pi$ for
harmonic 2 (the Nyquist frequency). Similar models, with Gaussian
responses, are described in West and Harrison (1997), and Harvey
(2004). The covariance matrix $\Omega$ of $\omega_t$ is set as the
block diagonal matrix $\Omega=\textrm{block
diag}(\Omega_1,\Omega_2)$, where $\Omega_1=1000 I_2$ corresponds to
the linear trend component, $\Omega_2=100I_3$ corresponds to the
seasonal component and it is chosen so that the trend has more
variability than the seasonal component (West and Harrison, 1997).
The priors $m_0$ and $P_0$ are set as $m_0=[0~0~0~0~0]'$ and
$P_0=1000I_5$, suggesting a weakly informative prior specification.
Figure \ref{fig1} plots the one-step forecast mean of $\{y_t\}$
against $\{y_t\}$. We see that the forecasts fit the data very
closely proposing a good model fit.

\subsubsection{Poisson}

In the context of generalized linear models, the Poisson
distribution (Johnson {\it et al.}, 2005) is associated with
modelling count data (Dobson, 2002). In a time series setting count
data are developed as in Jung {\it et al.} (2006).

Suppose that $\{y_t\}$ is a count time series, so that, for a
positive real-valued $\lambda_t>0$, $y_t|\lambda_t$ follows the
Poisson distribution, with density
$$
p(y_t|\lambda_t)=\textrm{exp}(-\lambda_t)\frac{\lambda_t^{y_t}}{y_t!},
\quad y_t=0,1,2,\ldots; \quad \lambda_t>0,
$$
where $y_t!$ denotes the factorial of $y_t$.

We can easily verify that this density is of the form (\ref{exp}),
with $z(y_t)=y_t$, $a(\phi_t)=\phi_t=1$, $\gamma_t=\log\lambda_t$,
$b(\gamma_t)=\textrm{exp}(\gamma_t)$, and $c(y_t,\phi_t)=1/y_t!$. We
can see that
$\E(y_t|\lambda)=\,db(\gamma_t)/\,d\gamma_t=\textrm{exp}(\gamma_t)=\lambda_t$
and
$\var(y_t|\lambda_t)=\,d^2b(\gamma_t)/\,\gamma_t^2=\textrm{exp}(\gamma_t)=\lambda_t$.

From the prior of $\gamma_t|y^{t-1}$ and the transformation
$\gamma_t=\log\lambda_t$, we obtain the prior of $\lambda_t|y^{t-1}$
as a gamma distribution, i.e. $\lambda_t|y^{t-1}\sim G(r_t,s_t)$,
with density
$$
p(\lambda_t|y^{t-1})=\frac{s_t^{r_t}}{\Gamma(r_t)} \lambda^{r_t-1}
\textrm{exp}(-s_t\lambda_t),
$$
for $r_t,s_t>0$. Then it follows that the posterior of $\lambda_t$
is the gamma $G(r_t+y_t,s_t+1)$.

For the definition of $r_t$ and $s_t$ we use the logarithmic link
$g(\lambda_t)=\log\lambda_t=\eta_t=F'\theta_t$ or
$\lambda_t=\textrm{exp}(F'\theta_t)$. Based on an evaluation of the
mean and variance of $\log\lambda_t$ and a numerical approximation
of the digamma function (see appendix), we can see
\begin{equation}\label{eq:poisson:rt}
r_t=\frac{1}{q_t} \quad \textrm{and} \quad
s_t=\frac{\exp(-f_t)}{q_t},
\end{equation}
where $f_t$ and $q_t$ are the mean and variance of $\eta_t$.

For the computation of $f_t^*$ and $q_t^*$, the posterior mean and
variance of $\gamma_t$, first define the digamma function $\psi(.)$
as $\psi(x)=\,d\log\Gamma(x)/\,dx$, where $\Gamma(.)$ denotes the
gamma function and of course $x>0$. Then we have
$$
f_t^*=\psi(r_t+y_t)-\log (s_t+1) \quad \textrm{and} \quad
q_t^*=\left.\frac{\,d\psi(x)}{\,dx}\right|_{x=r_t+y_t},
$$
which can be computed by the recursions $\psi(x)=\psi(x+1)-x^{-1}$
and $\,d\psi(x)/\,dx=\,d\psi(x+1)/\,dx+x^{-2}$. Using the
approximations $\psi(x)=\log x+(2x)^{-1}$ and
$\,d\psi(x)/\,dx=x^{-1}(1-(2x)^{-1})$, we can write
$$
f_t^*\approx \log \frac{r_t+y_t}{s_t+1}+\frac{1}{2(r_t+y_t)} \quad
\textrm{and} \quad q_t^*\approx\frac{2r_t+2y_t-1}{2(r_t+y_t)^2}.
$$
With $r_t$, $s_t$, $f_t^*$ and $q_t^*$ we can compute the first two
moments of $\theta_t|y^t$ as in (\ref{post:th1}). For a detailed
discussion on digamma functions the reader is referred to Abramowitz
and Stegun (1964, \S6.3).

Defining $r_t(\ell)$ and $s_t(\ell)$ according to $f_t(\ell)$ and
$q_t(\ell)$ and equation (\ref{eq:poisson:rt}), the $\ell$-step
forecast distribution of $y_{t+\ell}|y^t$ is given by
$$
p(y_{t+\ell}|y^t)=\binom{r_t(\ell)+y_{t+\ell}-1}{y_{t+\ell}} \left(
\frac{s_t(\ell)}{1+s_t(\ell)}\right)^{r_t(\ell)} \left(
\frac{1}{1+s_t(\ell)}\right)^{y_{t+\ell}},
$$
which is a negative binomial distribution. The forecast mean and
variance can be calculated by using conditional expectations, i.e.
$$
y_t(\ell)=\E(y_{t+\ell}|y^t)=\E(\E(y_{t+\ell}|\lambda_{t+\ell})|y^t)=
\frac{r_t(\ell)}{s_t(\ell)}
$$
and
$$
\var(y_{t+\ell}|y^t)=\E(\var(y_{t+\ell}|\lambda_{t+\ell})|y^t) +
\var(\E(y_{t+\ell}|\lambda_{t+\ell})|y^t) =
\frac{r_t(\ell)(s_t(\ell)+1)}{(s_t(\ell))^2}.
$$

The power discounting yields
$$
r_{t+1}=\delta (r_t+y_t)+1-\delta \quad \textrm{and} \quad
s_{t+1}=\delta(s_t+1).
$$

Considering the random walk evolution for $\theta_t$ so that
$\eta_t=\theta_t=\theta_{t-1}+\omega_t$, where $\omega_t\sim
N(0,\Omega)$, for some variance $\Omega$, we can see that
$$
\lambda_t=\textrm{exp}(\omega_t)\lambda_{t-1},
$$
since $\log\lambda_t=\eta_t=\theta_t$. Then from the normal
distribution of $\omega_t$, the distribution of
$\lambda_t|\lambda_{t-1}$ is
$$
p(\lambda_t|\lambda_{t-1}) = \frac{1}{\sqrt{2\pi\Omega}\lambda_t}
\textrm{exp}\left( -
\frac{(\log\lambda_t-\log\lambda_{t-1})^2}{2\Omega}\right),
$$
which is a log-normal distribution (see Section \ref{continuous}).
Tnen from (\ref{logl}) the log-likelihood function is
$$
\ell(\lambda_1,\ldots,\lambda_T;y^T) = \sum_{t=1}^T \left(
y_t\log\lambda_t -\lambda_t-\log y_t!
-\log\sqrt{2\pi\Omega}\lambda_t -
\frac{(\log\lambda_t-\log\lambda_{t-1})^2}{2\Omega}\right)
$$
Bayes factors can be calculated using (\ref{bf1}) and the negative
binomial one-step ahead forecast probability functions
$p(y_{t+1}|y^t)$.

\begin{figure}[t]
 \epsfig{file=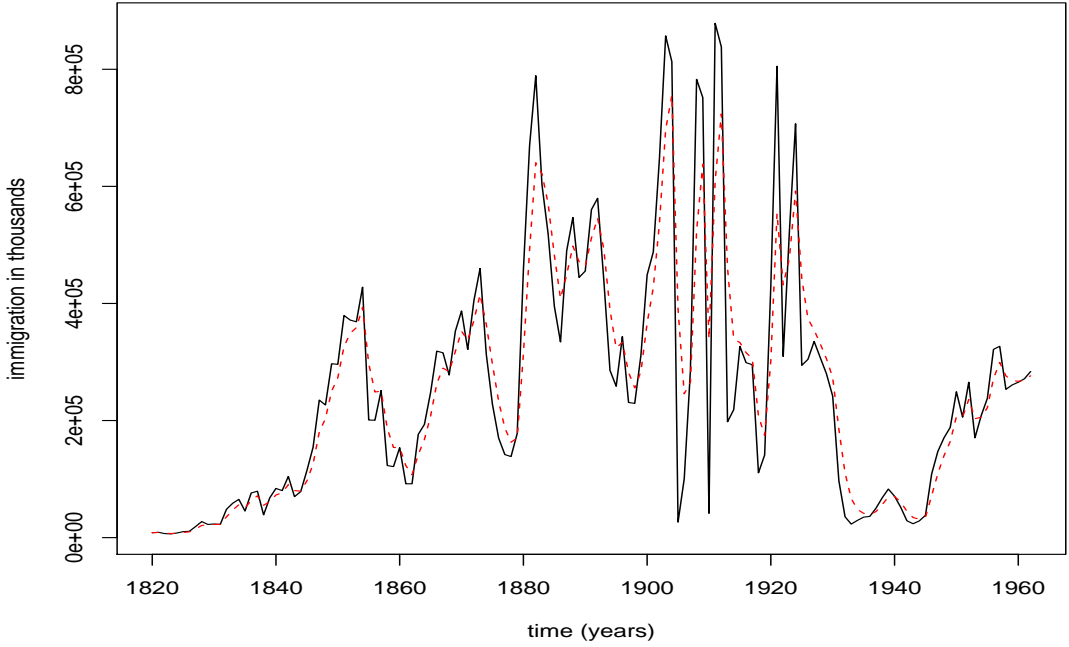, height=10cm, width=15cm}
 \caption{US annual immigration in thousands (solid line) and
 one-step forecast mean (dashed line).}\label{fig2}
\end{figure}

In order to illustrate the Poisson model we consider US annual
immigration data, in the period of 1820 to 1960. The data, which are
described in Kendall and Ord (1990, page 13), are shown in Figure
\ref{fig2}. The nature of the data fits to the assumption of a
Poisson distribution, but it can be argued that, after applying a
suitable transformation, some Gaussian time series model can be
appropriate. The data are non-stationary and a visual inspection
shows that they exhibit a local level behaviour. One simple model to
consider is the random walk evolution of $\eta_t=\theta_t$ as
described above. We use power discounting with $\delta=0.5$, which
is a low discount factor capable to capture the peak values of the
data. Figure \ref{fig2} shows the one-step forecast mean against the
actual data and as we see the forecasts capture well the immigration
data.

\subsubsection{Negative binomial and geometric}

The negative binomial distribution (Johnson {\it et al.}, 2005)
arises in many practical situations and it can be generated via
independent Bernoulli trails or via the Poisson/gamma mixture. In
time series analysis, an application of negative binomial responses
is given in Houseman {\it et al.} (2006). We note that the negative
binomial distribution includes the geometric as a special case (see
below).

Suppose that the time series $\{y_t\}$ is generated from the
negative binomial distribution, with probability function
$$
p(y_t|\pi_t)=\binom{y_t+n_t-1}{n_t-1}\pi_t^{n_t}(1-\pi_t)^{y_t},
\quad y_t=0,1,2,\ldots; \quad 0<\pi_t<1,
$$
where $\pi_t$ is the probability of success and $n_t$ is the number
of successes. This distribution belongs to the exponential family
(\ref{exp}), with $z(y_t)=y_t$, $a(\phi_t)=\phi_t=1$,
$\gamma_t=\log(1-\pi_t)$,
$b(\gamma_t)=-n_t\log(1-\textrm{exp}(\gamma_t))$, and
$c(y_t,\phi_t)=\binom{y_t+n_t-1}{n_t-1}$. Then it follows that
$\E(y_t|\pi_t)=\,db(\gamma_t)/\,d\gamma_t=n_t(1-\pi_t)/\pi_t$ and
$\var(y_t|\pi_t)=\,d^2b(\gamma_t)/\,d\gamma_t^2=n_t(1-\pi_t)/\pi_t^2$.
We note that by setting $n_t=1$ and $x_t=y_t-1$, the time series
$x_t$ follows a geometric distribution and thus all what follows
applies readily to the geometric distribution too.

By using the prior of $\gamma_t|y^{t-1}$ and the transformation
$\gamma_t=\log(1-\pi_t)$, the prior of $\pi_t|y^{t-1}$ is the beta
distribution $\pi_t|y^{t-1}\sim B(n_ts_t+1,r_t)$ and the posterior
of $\pi_t|y^t$ is the beta $\pi_t|y^t\sim B(n_ts_t+n_t+1,r_t+y_t)$.
Using the logit link, as in the binomial example, the definitions of
$r_t$ and $s_t$ are
$$
r_t=\frac{1+\exp(-f_t)}{q_t}\quad\textrm{and}\quad
s_t=\frac{1+\exp(f_t)-q_t}{n_tq_t}
$$
and the posterior moments $f_t^*$ and $q_t^*$ are
$$
f_t^*=\psi(n_ts_t+n_t+1)-\psi(r_t+y_t) \quad \textrm{and} \quad
q_t^*=\left.\frac{\,d\psi(x)}{\,dx}\right|_{x=n_ts_t+n_t+1} +
\left.\frac{\,d\psi(x)}{\,dx}\right|_{x=r_t+y_t},
$$
which can be approximated by
$$
f_t^*\approx\log \frac{n_ts_t+n_t+1}{r_t+y_t} +
\frac{1}{2(n_ts_t+n_t+1)}-\frac{1}{2(r_t+y_t)}
$$
and
$$
q_t^*\approx
\frac{2n_ts_t+2n_t+1}{2(n_ts_t+n_t+1)^2}+\frac{2r_t+2y_t-1}{2(r_t+y_t)^2}.
$$
Thus we can compute the moments of $\theta_t|y^t$ as in
(\ref{post:th1}) and so we obtain an approximation of the quantities
$r_t(\ell)$ and $s_t(\ell)$, as functions of $f_t(\ell)$ and
$q_t(\ell)$.

The $\ell$-step forecast distribution is given by
$$
p(y_{t+\ell}|y^t) = \frac{\Gamma(r_t(\ell)+n_{t+\ell}+s_t(\ell)+1)
\Gamma(r_t(\ell)+y_{t+\ell})
\Gamma(n_{t+\ell}s_t(\ell)+n_{t+\ell}+1) }{ \Gamma(r_t(\ell))
\Gamma(n_{t+\ell}s_t(\ell)+1)
\Gamma(r_t(\ell)+y_{t+\ell}+n_{t+\ell}s_t(\ell)+n_{t+\ell}+1) }
\binom{y_{t+\ell}+n_{t+\ell}-1}{n_{t+\ell}-1}.
$$
The forecast mean and variance of $y_{t+\ell}$ are given by
$$
y_t(\ell)=\E(y_{t+\ell}|y^t)=\E(\E(y_{t+\ell}|\pi_{t+\ell})|y^t)=\frac{r_t(\ell)}{s_t(\ell)}
$$
and
\begin{eqnarray*}
\var(y_{t+\ell}|y^t)&=&\E(\var(y_{t+\ell}|\lambda_{t+\ell})|y^t) +
\var(\E(y_{t+\ell}|\lambda_{t+\ell})|y^t) \\ &=&
\frac{(r_t(\ell)+n_{t+\ell}s_t(\ell))(r_t(\ell)+n_{t+\ell}r_t(\ell)+n_{t+\ell}^2
s_t(\ell)-n_{t+\ell})}{s_t(\ell)(n_{t+\ell}s_t(\ell)-1)} -
\frac{r_t(\ell)^2}{n_{t+\ell}^2s_t(\ell)^2}.
\end{eqnarray*}

The power discounting yields
$$
r_{t+1}=\delta(r_t+y_t-1)+1 \quad \textrm{and} \quad
s_{t+1}=\frac{\delta(n_ts_t+n_t)}{n_{t+1}},
$$
where as usual $\delta$ is a discount factor.

Considering the random walk evolution for
$\eta_t=\theta_t=\theta_{t-1}+\omega_t$, the link $\log
n_t(1-\pi_t)/n_t=\eta_t$, yields the evolution for $\pi_t$
\begin{equation}\label{nb:pi}
\pi_t=\frac{\pi_{t-1}}{\pi_{t-1}+\exp(\omega_t)-\pi_{t-1}\exp(\omega_t)}.
\end{equation}
Given that $\omega_t\sim N(0,\Omega)$, for a known variance
$\Omega$, the distribution of $\pi_t|\pi_{t-1}$ is
$$
p(\pi_t|\pi_{t-1})=\frac{1}{\sqrt{2\pi\Omega}\pi_t(1-\pi_t)}
\exp\left(-\frac{1}{2\Omega}\left(\log\frac{\pi_{t-1}(1-\pi_t)}{\pi_t(1-\pi_{t-1})
}\right)^2\right)
$$
and so from (\ref{logl}) the log-likelihood function is
\begin{eqnarray*}
\ell(\pi_1,\ldots,\pi_T;y^T) &=& \sum_{t=1}^T \bigg(
y_t\log(1-\pi_t) +n_t\log\pi_t + \log \binom {y_t+n_t-1}{n_t-1} \\
&& -\log\sqrt{2\pi\Omega}\pi_t(1-\pi_t) -\frac{1}{2\Omega}\left(
\log\frac{\pi_{t-1}(1-\pi_t)}{\pi_t(1-\pi_{t-1})}\right)^2\bigg)
\end{eqnarray*}
Bayes factors can be computed using (\ref{bf1}) and the predictive
distribution $p(y_{t+1}|y^t)$.

\begin{figure}[t]
 \epsfig{file=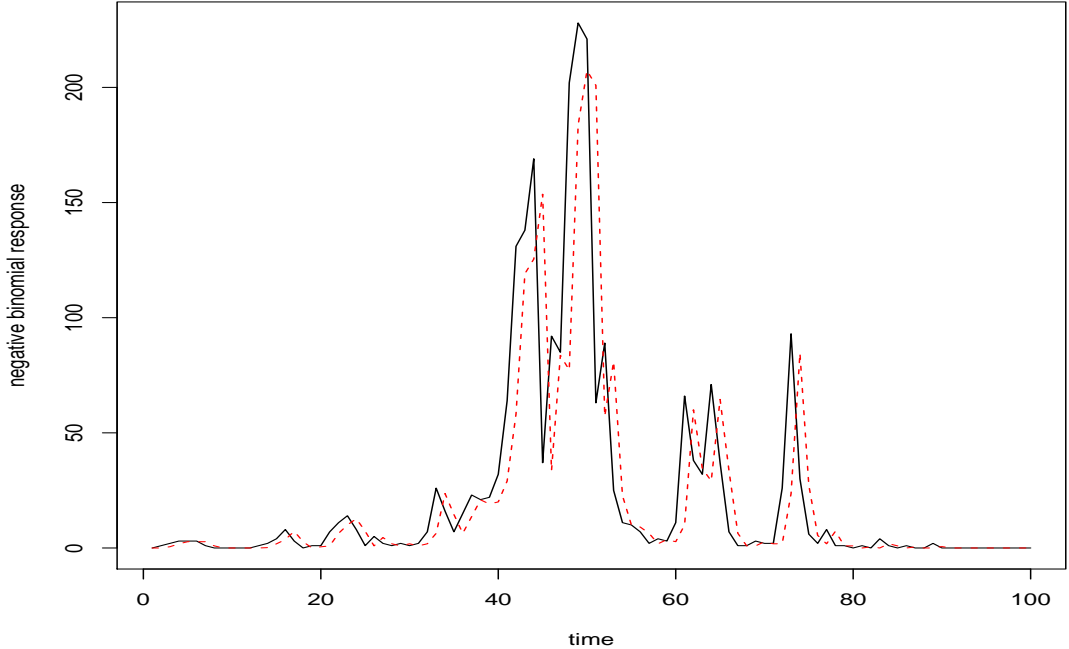, height=10cm, width=15cm}
 \caption{Negative binomial simulated data (solid line) and
 one-step forecast mean (dashed line).}\label{fig2a}
\end{figure}

To illustrate the above model we have simulated 100 observations
from the above model; we simulate one draw from $\pi_0\sim B(2,1)$
so that $\E(\pi_0)=2$, we simulate 100 innovations
$\omega_1,\ldots,\omega_{100}$ from a $N(0,1)$, then using
(\ref{nb:pi}) we generate $\pi_1,\ldots,\pi_{100}$ and finally, for
each time $t$, we simulate one draw from a negative binomial with
parameters $n_t=n=10$ and $\pi_t$. Figure \ref{fig2a} shows the
simulated data (solid line) together with the one-step ahead
forecast means $r_t/s_t$. For the fit, we pretend we did not have
knowledge of the simulation process and so we have specified
$F=[1~0]'$, $G=\Omega=I_2$ (the $2\times 2$ identity matrix),
$m_0=[0~0]'$, and $P_0=1000I_2$, the last indicating a weakly
informative prior specification (i.e. $P_0^{-1}\approx 0$). We
observe that the forecasts follow the data closely indicating a good
fit. We have found that as it is well known for Gaussian time
series, these prior settings are insensitive to forecasts, since
prior information is deflated with time.

\subsection{Continuous distributions for the response $y_t$}\label{continuous}

\subsubsection{Normal}

Normal or Gaussian time series are discussed extensively in the
literature, see e.g. West and Harrison (1997) for a Bayesian
treatment of Gaussian state-space models. Here we discuss Gaussian
responses in the DGLM setting, for completeness purposes, but also
because the normal distribution has many similarities with the
log-normal distribution that follows.

Suppose that $\{y_t\}$ is a time series generated from a normal
distribution, i.e. $y_t|\mu_t\sim N(\mu_t,V)$, with density
$$
p(y_t|\mu_t)=\frac{1}{\sqrt{2\pi V}}
\exp\left(-\frac{(y_t-\mu_t)^2}{2V}\right), \quad
-\infty<y_t,\mu_t<\infty; \quad V>0,
$$
where $\mu_t$ is the level of $y_t$. The variance $V$ of the process
can be time-varying, but for simplicity here, we assume it
time-invariant. Here, this variance is assumed known, while $\mu_t$
is assumed unknown. If $V$ is unknown, Bayesian inference is
possible by assuming that $1/V$ follows a gamma distribution and
this model leads to a conjugate analysis (resulting to a posterior
gamma distribution for $1/V$ and to a Student $t$ distribution for
the forecast distribution of $y_{t+\ell}$). This model is examined
in detail in West and Harrison (1997, Chapter 4). Returning to the
above normal density, we can easily see that $p(y_t|\mu_t)$ is of
the form of (\ref{exp}), with $z(y_t)=y_t$,
$a(\phi_t)=\phi_t^{-1}=V$, $\gamma_t=\mu_t$,
$b(\gamma_t)=\gamma_t^2/2$ and $c(y_t,\phi_t)=(2\pi
V)^{-1/2}\exp(-y_t^2/(2V)$.

The prior for $\mu_t|y^{t-1}$ is the normal distribution
$\mu_t|y^{t-1}\sim N(r_ts_t^{-1},s_t^{-1})$ and the posterior of
$\mu_t|y^t$ is the normal distribution
$$
\mu_t|y^t\sim
N\left(\frac{r_t+V^{-1}y_t}{s_t+V^{-1}},\frac{1}{s_t+V^{-1}}\right).
$$
The link function is the identity link, i.e. $g(\mu_t)=\mu_t$ and so
we have $\mu_t=\eta_t=F'\theta_t$, which implies $r_t=f_t/q_t$ and
$s_t=1/q_t$. By replacing these quantities in the above prior and
posterior densities, we can verify the Kalman filter recursions.

It turns out that the $\ell$-step forecast distribution is also a
normal distribution, i.e.
$$
y_{t+\ell}|y^t \sim N\left(
\frac{r_t(\ell)}{s_t(\ell)},V+\frac{1}{s_t(\ell)}\right).
$$

The power discounting yields
$$
r_{t+1}=\delta^2(r_t+V^{-1}y_t) \quad \textrm{and} \quad
s_{t+1}=\delta^2(s_t+V^{-1}).
$$

Adopting the random walk evolution for
$\theta_t=\theta_{t-1}+\omega_t$, from the identity link
$\mu_t=\eta_t=\theta_t$, we have that $\mu_t|\mu_{t-1}\sim
N(\mu_{t-1},\Omega)$, where $\omega_t\sim N(0,\Omega)$. From
(\ref{logl}) the log-likelihood function is
$$
\ell(\mu_1,\ldots,\mu_T;y^T)=\sum_{t=1}^T \left(
\frac{1}{2V}(2y_t\mu_t-\mu_t^2)
-\log\sqrt{4\pi^2V\Omega}-\frac{y_t^2}{2V}-\log\frac{(\mu_t-\mu_{t-1})^2}{2\Omega}\right).
$$
Bayes factors can be easily computed from the forecast density
$p(y_{t+1}|y^t)$ and the Bayes factor formula (\ref{bf1}).

\subsubsection{Log-normal}

The log-normal distribution has many applications, e.g. in
statistics (Johnson {\it et al.}, 1994), in economics (Aitchison and
Brown, 1957), and in life sciences (Limpert {\it et al.}, 2001).

Suppose that the time series $\{y_t\}$ is generated from a
log-normal distribution, with density
$$
p(y_t|\lambda_t)=\frac{1}{\sqrt{2\pi V}} \exp\left(-\frac{(\log
y_t-\lambda_t)^2}{2V}\right),\quad y_t>0; \quad -\infty
<\lambda_t<\infty; \quad V>0,
$$
where $\log y_t | \lambda_t \sim N(\lambda_t,V)$. We will write
$y_t|\lambda_t\sim LogN(\lambda_t,V)$. This distribution is of the
form of (\ref{exp}), with $z(y_t)=\log y_t$,
$a(\phi_t)=\phi_t^{-1}=V$, $\gamma_t=\lambda_t$,
$b(\gamma_t)=\gamma_t^2/2$ and $c(y_t,\phi_t)=(2\pi
V)^{-1/2}y_t^{-1}\exp(-(\log y_t)^2/(2V))$.

From the normal part we can see
$$
\E(\log
y_t|\lambda_t)=\frac{\,db(\gamma_t)}{\,d\gamma_t}=\lambda_t
$$
and from the log-normal part we can see
$$
\E(y_t|\lambda_t)=\exp(\lambda_t+V/2)=\mu_t
$$
from the latter of which the logarithmic link can be suggested,
i.e. $\eta_t=\log\mu_t=\lambda_t+V/2$.

From the normal distribution of $\log y_t$, it follows that the
prior distribution of $\lambda_t|y^{t-1}$ is
$$
\lambda_t|y^{t-1}\sim N\left(\frac{r_t}{s_t},\frac{1}{s_t}\right)
$$
and the posterior distribution of $\lambda_t|y^t$ is
$$
\lambda_t|y^t\sim N\left(\frac{r_t+V^{-1}\log y_t}{s_t+V^{-1}},
\frac{1}{s_t+V^{-1}}\right),
$$
where $r_t$ and $s_t$ are calculated as in the normal case, i.e.
$r_t=f_t/q_t$ and $s_t=1/q_t$. With the definitions of $r_t(\ell)$
and $s_t(\ell)$, we have that the $\ell$-step forecast
distribution of $y_{t+\ell}$ is
$$
y_{t+\ell}|y^t\sim
LogN\left(\frac{r_t(\ell)}{s_t(\ell)},V+\frac{1}{s_t(\ell)}\right).
$$
The forecast mean of $y_{t+\ell}$ is
$$
y_t(\ell)=\E(y_{t+\ell}|y^t)=\exp\left(\frac{r_t(\ell)}{s_t(\ell)}+\frac{1}{2s_t(\ell)}
\right)
\exp\left(\frac{V}{2}\right)=\exp\left(\frac{2f_t(\ell)+q_t(\ell)+V}{2}\right),
$$
where $f_t(\ell)$ and $q_t(\ell)$ are the respective mean and
variance of $\eta_{t+\ell}$, given information $y^t$.

Considering power discounting, the updating of $r_t$ and $s_t$ is
$$
r_{t+1}=\delta^2(r_t+V^{-1}\log y_t) \quad \textrm{and} \quad
s_{t+1}=\delta^2(s_t+V^{-1}).
$$

Adopting the random walk evolution for
$\eta_t=\theta_t=\theta_{t-1}+\omega_t$, the distribution of
$\lambda_t|\lambda_{t-1}$ is normal, i.e.
$\lambda_t|\lambda_{t-1}\sim N(\lambda_{t-1},\Omega)$, where
$\Omega$ is the variance of $\omega_t$. From (\ref{logl}) the
log-likelihood function is obtained as
\begin{eqnarray*}
\ell(\lambda_1,\ldots,\lambda_T;y^T)&=&\sum_{t=1}^T
\bigg(\frac{1}{2V}(2\lambda_t\log y_t-\lambda_t^2) -\log
\sqrt{4\pi^2V\Omega} -\log y_t \\ && - \frac{(\log y_t)^2}{2V} -
\log\frac{(\lambda_t-\lambda_{t-1})^2}{2\Omega} \bigg).
\end{eqnarray*}
Bayes factors can be calculated from (\ref{bf1}) and the log-normal
predictive density $p(y_{t+1}|y^t)$. As an example, consider the
comparison of two models $\mathcal{M}_1$ and $\mathcal{M}_2$, which
differ in the variances $V_1$ and $V_2$, respectively. Then, by
denoting $r_{1t}$, $s_{1t}$, $r_{2t}$ and $s_{2t}$, the values of
$r_t$, $s_t$, for $\mathcal{M}_j$ $(j=1,2)$, we can express the
logarithm of the Bayes factor $H_t(1)$ as
$$
\log H_t(1) = \frac{1}{2}\log
\frac{V_2+s_2,{t+1}^{-1}}{V_1+s_{1,t+1}^{-1}} + \frac{ (\log y_{t+1}
- r_{2,t+1}s_{2,t+1}^{-1})^2}{2(V_2-s_{2,t+1}^{-1})} - \frac{ (\log
y_{t+1} - r_{1,t+1}s_{1,t+1}^{-1})^2}{2(V_1-s_{1,t+1}^{-1})}.
$$
By comparing $\log H_t(1)$ to 0, we can conclude preference of
$\mathcal{M}_1$ or $\mathcal{M}_2$, i.e. if $\log H_t(1)>0$ we
favour $\mathcal{M}_1$, if $\log H_t(1)<0$ we favour
$\mathcal{M}_1$, while if $\log H_t(1)=0$ the two models are
equivalent, in the sense that they both produce the same one-step
forecast distributions.

\begin{table}
\caption{Mean square error (MSE) and Log-likelihood function $(\ell(
.))$ for several values of the discount factor $\delta$ for the
log-normal data.}\label{table:logn}
\begin{center}
\begin{tabular}{|c||ccccccccc|}
\hline $\delta$ & 0.2 & 0.3 & 0.4 & 0.5 & 0.6 & 0.7 & 0.8 & 0.9 &
0.99 \\ $MSE$ & 103.75 & 13.34 & 3.16 & \textbf{2.22} & 2.72 & 3.37 & 3.93 & 4.34 & 4.57 \\
$\ell(.)$ & -35.26 & -35.28 & -35.34 & -35.44 & -35.61 & -35.86 &
-36.2 & -36.60 & -36.93 \\ \hline
\end{tabular}
\end{center}
\end{table}

\begin{figure}[h]
 \epsfig{file=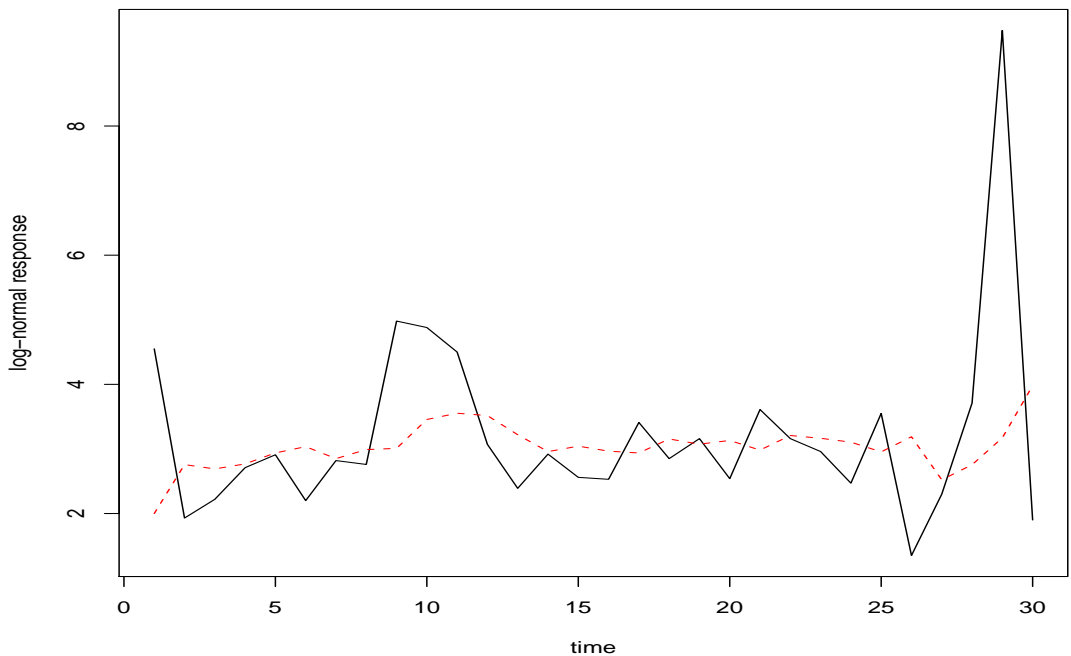, height=10cm, width=15cm}
 \caption{Log-normal data (solid line) and one-step forecasts (dotted line) for $\delta=0.5$.}\label{fig:logn}
\end{figure}

To illustrate the above DGLM for log-normal data we consider
production data, consisting of 30 consecutive values of value of a
product; these data are reported in Morrison (1958). A simple
histogram shows that these data are positively skewed and it can be
argued that the data exhibit local level time series dependence.
Morrison (1958) show that modelling these data with the normal
distribution can lead to inappropriate control. Here we use the
power discounting approach to update $r_t$ and $s_t$; Table
\ref{table:logn} shows the mean square forecast error (MSE) and the
value of the log-likelihood function evaluated at the posterior mean
$\E(\lambda_t|y^t)$ for a range of values of $\delta$. The result is
that $\delta=0.5$ produces the smallest MSE, while the likelihood
function does not change dramatically. Figure \ref{fig:logn} plots
the one-step forecasts for $\delta=0.5$ against the actual data.
Although the extreme value $y_{29}=9.48$ is poorly predicted, we
conclude that the overall forecast performance of this model is
good, especially given the short length of this time series.

\subsubsection{Gamma}

The gamma distribution (Johnson {\it et al.}, 1994) is perhaps one
of the most used continuous distributions, as it can serve as a
model for the variance or precision of a population or experiment.
In particular in Bayesian inference it is a very popular choice as
the conjugate prior for the inverse of the variance of a linear
conditionally Gaussian model (see also the discussion of the normal
distribution above).

Suppose that $\{y_t\}$ is a time series generated from a gamma
distribution, with density
$$
p(y_t|\alpha_t,\beta_t)=\frac{\beta_t^{\alpha_t}}{\Gamma(\alpha_t)}
y_t^{\alpha_t-1} \exp(-\beta_t y_t), \quad y_t>0; \quad
\alpha_t,\beta_t>0.
$$
This distribution is referred to as $y_t|\alpha_t,\beta_t\sim
G(\alpha_t,\beta_t)$. Our interest is focused on $\beta_t$ and so we
will assume that $\alpha_t$ is known {\it a priori}. Thus we write
$p(y_t|\alpha_t,\beta_t)\equiv p(y_t|\beta_t)$.

The above gamma distribution is of the form of (\ref{exp}), with
$z(y_t)=y_t$, $a(\phi_t)=\phi_t=1$, $\gamma_t=-\beta_t$,
$b(\gamma_t)=-\log ((-\gamma_t)^{\alpha_t}/\Gamma(\alpha_t))$ and
$c(y_t,\phi_t)=y_t^{\alpha_t-1}$.

It follows that
$$
\E(y_t|\beta_t)=\frac{\,db(\gamma_t)}{\,d\gamma_t}=\frac{\alpha_t}
{\beta_t}=\mu_t>0
$$
and
$$
\var(y_t|\beta_t)=\frac{\,d^2b(\gamma_t)}{\,d\gamma_t^2}=\frac{\alpha_t}
{\beta_t^2}.
$$

The prior and posterior distributions of $\beta_t$ are gamma, i.e.
$\beta_t|y^{t-1}\sim G(\alpha_ts_t+1,r_t)$ and $\beta_t|y^t\sim
G(\alpha_ts_t+\alpha_t+1,r_t+y_t)$.

Since $\mu_t>0$, the logarithmic link is a appropriate, i.e.
$g(\mu_t)=\log\mu_t=\eta_t=F'\theta_t$. Then $r_t$ and $s_t$ are
defined in a similar way as in the Poisson case, i.e.
$$
r_t=\frac{\exp(-f_t)}{q_t}\quad \textrm{and} \quad
s_t=\frac{1-q_t}{\alpha_tq_t},
$$
where $\alpha_ts_t+1>0$. The posterior moments of $\log\mu_t$ are
given by
$$
f_t^*=\psi(\alpha_ts_t+y_t+1)-\log(r_t+1) \quad \textrm{and} \quad
q_t^*=\left.\frac{\,d\psi(x)}{\,dx}\right|_{x=\alpha_ts_t+y_t+1},
$$
which can be approximated, as in the Poisson case, by
$$
f_t^*\approx \log \frac{\alpha_ts_t+y_t+1}{r_t+1} +
\frac{1}{2(\alpha_ts_t+y_t+1)} \quad \textrm{and} \quad q_t^*
\approx \frac{2\alpha_ts_t+2y_t+1}{2(\alpha_ts_t+y_t+1)}.
$$

With the definition of $r_t(\ell)$ and $s_t(\ell)$, the
$\ell$-step forecast distribution is
$$
p(y_{t+\ell}|y^t)=\frac{r_t(\ell)^{\alpha_{t+\ell}s_t(\ell)}
\Gamma(\alpha_{t+\ell}s_t(\ell)+\alpha_{t+\ell}+1) }{
\Gamma(r_t(\ell)) \Gamma(\alpha_{t+\ell})}
y_{t+\ell}^{\alpha_{t+\ell}-1}(r_t(\ell)+y_{t+\ell})^{-(
\alpha_{t+\ell}s_t(\ell)+\alpha_{t+\ell}+1)}.
$$
The mean and variance of this distribution can be obtained by
conditional expectations, i.e
$$
y_t(\ell)=\E(y_{t+\ell}|y^t)=\E(\E(y_{t+\ell}|\beta_{t+\ell})|y^t)=\frac{r_t(\ell)}{
s_t(\ell)}
$$
and
$$
\var(y_{t+\ell}|y^t) = \E(\var(y_{t+\ell}|\beta_{t+\ell})|y^t)+
\var(\E(y_{t+\ell}|\beta_{t+\ell})|y^t) =
\frac{r_t(\ell)^2(s_t(\ell)+1)}{s_t(\ell)^2(\alpha_{t+\ell}s_t(\ell)
-1)}.
$$

The power discounting yields
$$
r_{t+1}=\delta(r_t+y_t) \quad \textrm{and} \quad
s_{t+1}=\frac{\delta \alpha_ts_t+\delta \alpha_t}{\alpha_{t+1}}.
$$

From the logarithmic link function we have
$\beta_t=\alpha_t/\exp(\eta_t)$ and if we consider a random walk
evolution for $\eta_t=\theta_t=\theta_{t-1}+\omega_t$, we obtain
the evolution of $\beta_t$ as
$$
\beta_t=\frac{\alpha_t\beta_{t-1}}{\alpha_{t-1}\exp(\omega_t)},
$$
which together with the normal distribution of $\omega_t\sim
N(0,\Omega)$, results to the distribution
$$
p(\beta_t|\beta_{t-1})=\frac{1}{\sqrt{2\pi \Omega}\beta_t}
\exp\left(-\frac{(\log\beta_t- \alpha_t\alpha_{t-1}^{-1}
\log\beta_{t-1})^2}{2\Omega}\right),
$$
which is the log-normal distribution $\beta_t|\beta_{t-1}\sim LogN
(\alpha_t\alpha_{t-1}^{-1}\log\beta_{t-1},\Omega)$. Note that the
above expressions can be simplified when $\alpha_t=\alpha$ is
time-invariant. Model comparison and model monitoring can be
conducted by considering the Bayes factors, which can be computed
from (\ref{bf1}) and the predictive density $p(y_{t+\ell}|y^t)$.

Bayes factors can be computed using (\ref{bf1}) and the predictive
distribution $p(y_{t+1}|y^t)$. Here we give two examples, both of
which are using the power discounting approach. In the first we
consider two competing models $\mathcal{M}_1$ and $\mathcal{M}_2$,
which differ in the discount factors $\delta_1$ and $\delta_2$,
respectively, but otherwise they have the same structure. Then, if
we denote $r_{it}$ and $s_{it}$ the values of $r_t$ and $s_t$ for
model $\mathcal{M}_i$ $(i=1,2)$, then the Bayes factor $H_t(1)$ can
be expressed as
$$
H_t(1)=\frac{ r_{1,t+1}^{\alpha s_{1,t+1}} \Gamma(\alpha
s_{1,t+1}+\alpha+1) (r_{1,t+1}+y_{t+1})^{-(\alpha
s_{1,t+1}+\alpha+1)} \Gamma(r_{2,t+1}) }{ r_{2,t+1}^{\alpha
s_{2,t+1}} \Gamma(\alpha s_{2,t+1}+\alpha+1)
(r_{2,t+1}+y_{t+1})^{-(\alpha s_{2,t+1}+\alpha+1)} \Gamma(r_{1,t+1})
},
$$
where, for simplicity we assume that $\alpha_t=\alpha$ is invariant
over time and known.

In the second example we consider a fixed discount factor
$\delta_1=\delta_2=\delta$, but now the two models $\mathcal{M}_1$
and $\mathcal{M}_2$ differ in the values of $\alpha$, namely
$\alpha_1$ and $\alpha_2$. Then we can see that
$r_t=r_{it}=\delta(r_{t-1}+y_{t-1})$ and $s_t=s_{it}=(\delta\alpha
s_{i,t-1}+\delta\alpha)/\alpha=\delta s_{t-1}+\delta$, since $r_t$
and $s_t$ do not depend on $\alpha_i$ (note that this would not be
the case if $\alpha_i$ were time-varying). Then the Bayes factor of
$\mathcal{M}_1$ against $\mathcal{M}_2$ can be expressed as
$$
H_t(1)=r_{t+1}^{s_{t+1}(\alpha_1-\alpha_2)}
y_{t+1}^{\alpha_1-\alpha_2}
(r_{t+1}+y_{t+1})^{(s_{t+1}+1)(\alpha_2-\alpha_1)} \frac{
\Gamma(\alpha_2) \Gamma(\alpha_1 s_{t+1}+\alpha_1+1)}{
\Gamma(\alpha_1) \Gamma(\alpha_2 s_{t+2}+\alpha_2+1)}.
$$
Thus, by comparing $H_t(1)$ with 1, we have a means for choosing the
parameter $\alpha$.

To illustrate the gamma distribution we give an example from
finance. Suppose that $y_t$ represents the continually compound
return, known also as log-return, of the price of an asset, defined
as $y_t=\log p_t - \log p_{t-1}$, where $p_t$ is the price of the
asset at time $t=1,\ldots,T$. In volatility modelling, one wishes to
estimate the conditional variance $\sigma_t^2$ of $y_t$. This plays
an important role in risk management and in investment strategies
(Chong, 2004), as it quantifies the uncertainty around assets. A
classical model is the generalized autoregressive heteroscedastic
(GARCH), which assumes that given $\sigma_t$, $y_t$ follows a normal
distribution, i.e. $y_t|\sigma_t\sim N(0,\sigma_t^2)$ and then it
specifies the evolution of $\sigma_t^2$ as a linear function of past
values of $\sigma_t^2$ and $y_t^2$. GARCH models are discussed in
detail in Tsay (2002).

From $y_t|\sigma_t\sim N(0,\sigma_t^2)$, we can see that, given
$\sigma_t$, $y_t^2/\sigma_t^2$ follows a chi-square distribution
with 1 degree of freedom or a $G(1/2,1/2)$. Thus $y_t^2|\sigma_t
\sim \sigma_t^2 G(1/2,1/2) \equiv G(1/2,1/(2\sigma_t^2))$. Then by
defining $\alpha_t=1/2$ and $\beta_t=1/(2\sigma_t^2)$, we have that
$y_t^2|\beta_t\sim G(1/2,\beta_t)$ and so we can apply the above
inference of the gamma response. Assuming a random walk evolution
for $\eta_t=\theta_t$, we have
$$
\beta_t=\frac{\beta_{t-1}}{\exp(\omega_t)} \Rightarrow
\sigma_t^2=\exp(\omega_t) \sigma_{t-1}^2,
$$
where $\omega_t$ is defined above.

We note that from power discounting we have $r_t=\delta
r_{t-1}+\delta y_{t-1}^2=\sum_{i=1}^{t-1}\delta^iy_{t-i}^2$ and
$s_t=\delta
s_{t-1}+\delta=\sum_{i=1}^{t-1}\delta^i=\delta(1-\delta^t)/(1-\delta)$.
Thus the one-step forecast mean of $y_t^2$ is
$$
\E(y_t^2|y^{t-1})=\frac{r_{t-1}(1)}{s_{t-1}(1)}=\frac{r_t}{s_t}=\frac{1-\delta}{\delta(1-\delta^t)}
\sum_{i=1}^{t-1} \delta^i y_{t-i}^2.
$$
From the prior of $\beta_t|y^{t-1}$, we can see that $1/\sigma_t^2 |
y^{t-1} \sim G((s_t+3)/2,r_t/2)$ and so $\sigma_t^2|y^{t-1}$ follows
an inverted gamma distribution, i.e. $\sigma_t^2|y^{t-1} \sim
IG((s_t+3)/2,r_t/2)$. Similarly, we can see that the posterior
distribution of $1/\sigma_t^2$ and $\sigma_t^2$ are $1/\sigma_t^2 |
y^t \sim G((s_t+3)/2,(r_t+y_t)/2)$ and $\sigma_t^2|y^t\sim
IG((s_t+3)/2,(r_t+y_t)/2)$, respectively. From these distributions
we can easily report means, variances and quantiles, as required.

\begin{figure}
 \epsfig{file=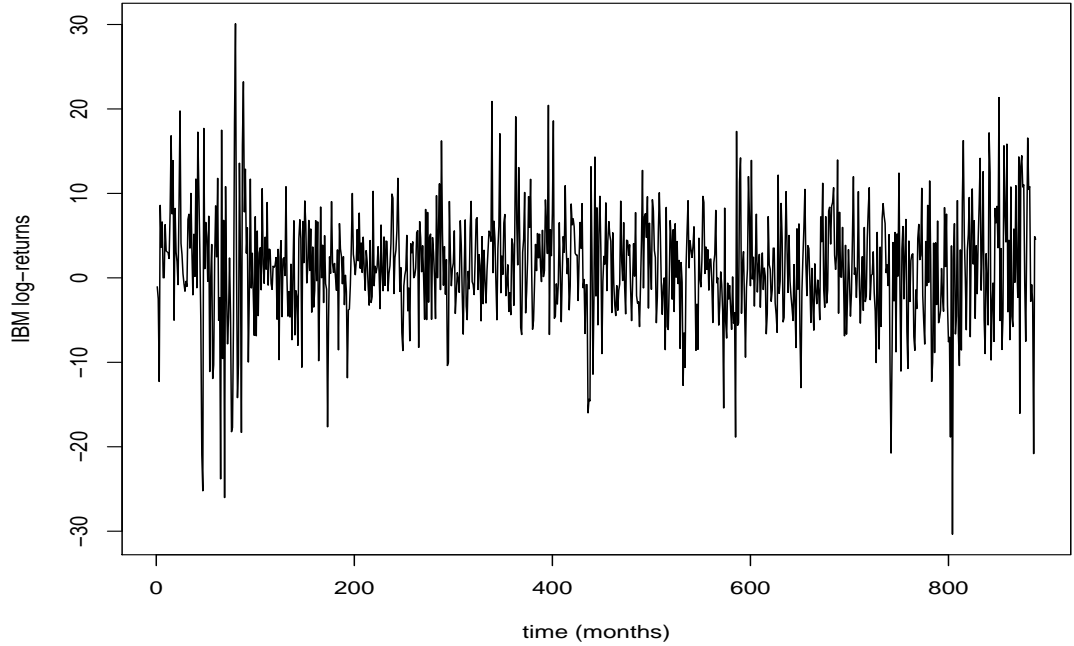, height=10cm, width=15cm}
 \caption{Log-returns of IBM stock prices.}\label{fig3}
\end{figure}

\begin{figure}
 \epsfig{file=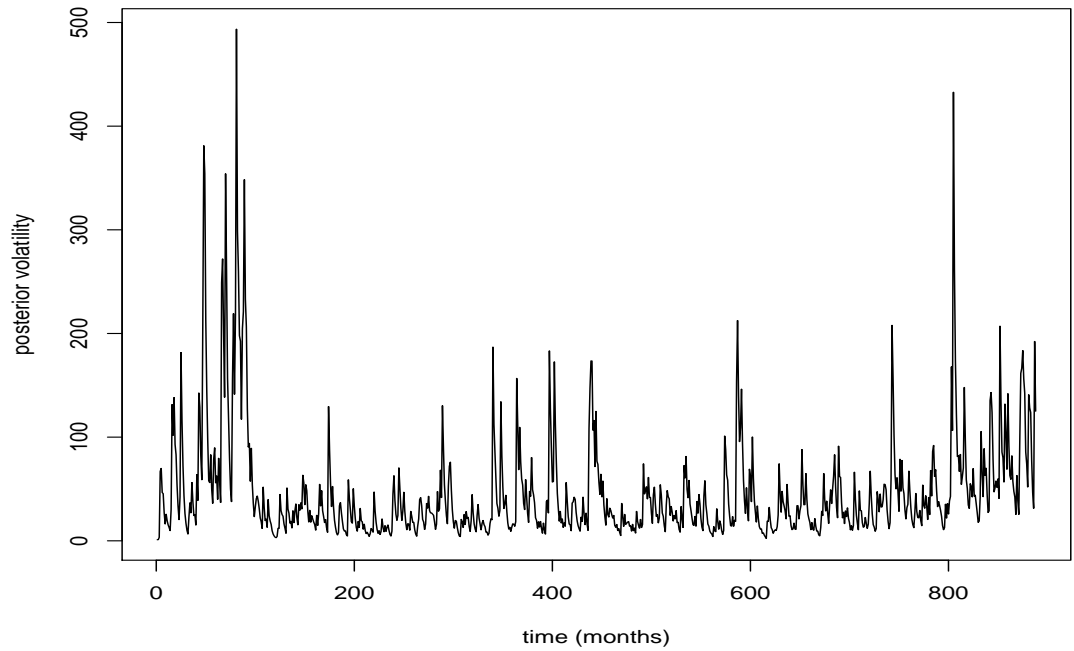, height=10cm, width=15cm}
 \caption{Posterior volatility of the IBM log-returns.}\label{fig4}
\end{figure}

We consider log returns from IBM stock prices over a period of 74
years. These data, which are described in Tsay (2002, Chapter 9),
are plotted in Figure \ref{fig3}. Figure \ref{fig4} shows the
posterior estimate of the volatility
$\widehat{\sigma}_t^2=\E(\sigma_t^2|y^t)$. We can see that the
volatile periods are captured well, e.g. the first 120 observations
in both figures indicate the high volatility. The model performance
can be assessed by looking at the log-likelihood function of
$\beta_t=1/(2\sigma_t^2)$, evaluated at the posterior mean
$\widehat{\sigma}_t^2$. The log-likelihood is
$$
\ell(\beta_1,\ldots,\beta_T;y^T) = -\frac{T}{2}\log (2\Omega\pi^2) -
\sum_{t=1}^T \log y_t^2 - \frac{1}{2\Omega} \sum_{t=1}^T (\log
\beta_t-\log \beta_{t-1})^2,
$$
where $\Omega$ is the variance of $\omega_t$ (the innovation of the
random walk evolution of $\eta_t=\theta_t$). Here $T=888$ and with
$\delta=0.6$ and $\Omega=100$, we compare this model with several
ARCH/GARCH models. Table \ref{table1} shows the log-likelihood
function of our model compared with those of the ARCH/GARCH. We see
that our model outperforms the ARCH/GARCH producing much larger
values of the log-likelihood function.

\begin{table}
\caption{Comparison of the gamma model with ARCH and GARCH models.
Shown are the log-likelihood functions of the models, using the IBM
data.}\label{table1}
\begin{center}
\begin{tabular}{|c||ccccc|}
\hline model & gamma & ARCH(1) & ARCH(2) & ARCH(3) & ARCH(4) \\
$\ell(.)$ & \textbf{-241.07} & -2133.79 & -2123.10 & -2115.11 & -2110.93 \\
model & & GARCH(1,1) & GARCH(1,2) & GARCH(2,1) & GARCH(2,2) \\
$\ell(.)$ & & -2109.33 & -2125.05 & -2130.86 & -2123.74
 \\
\hline
\end{tabular}
\end{center}
\end{table}

Inference and forecasting for the inverse or inverted gamma model is
very similar with the gamma model. For example suppose that given
$\alpha_t$ and $\beta_t$, the response $y_t$ follows the inverse
gamma distribution $y_t\sim IG(\alpha_t,\beta_t)$, so that
$$
p(y_t|\alpha_t,\beta_t)=\frac{\beta_t^{\alpha_t}}{\Gamma(\alpha_t)}
\frac{1}{y_t^{\alpha_t+1}} \exp\left(-\frac{\beta_t}{y_t}\right),
\quad y_t>0; \quad \alpha_t,\beta_t>0.
$$
Given $\alpha_t$ (as in the gamma case), the above inverse gamma
distribution is of the form of (\ref{exp}), with $z(y_t)=1/y_t$,
$a(\phi_t)=\phi_t=1$, $\gamma_t=-\beta_t$, $b(\gamma_t)=-\log
((-\gamma_t)^{\alpha_t}/\Gamma(\alpha_t))$ and
$c(y_t,\phi_t)=y_t^{-(\alpha_t+1)}$. The prior distribution for
$\beta_t$ is the gamma $\beta_t|y^{t-1}\sim G(\alpha_ts_t+1,r_t)$
and the posterior distribution is the gamma $\beta_t|y^t\sim
G(\alpha_ts_t+1,r_t+y_t^{-1})$. Thus the above prior is the same as
in the gamma model and the posterior changes slightly. As a result
inference and forecasting for the inverse gamma follows readily from
the gamma distribution.

\subsubsection{Weibull and exponential}

The exponential and the Weibull distributions can be used in
survival analysis, for example, in medicine, to estimate the
survival of patients, or in reliability, to estimate failure times
of say a manufacturing product. The exponential distribution is a
special case of the Weibull and for a discussion of both, the reader
is referred to Johnson {\it et al.} (1994).

Suppose that the time series $\{y_t\}$ is generated by a Weibull
distribution, with density function
$$
p(y_t|\lambda_t)=\frac{\nu_t}{\lambda_t}y_t^{\nu_t-1}\exp\left(-\frac{y_t^{\nu_t}}
{\lambda_t}\right), \quad y_t>0; \quad \lambda_t,\nu_t>0.
$$
Here we assume that $\nu_t$ is known and we note that for
$\nu_t=1$ we obtain the exponential distribution with parameter
$1/\lambda_t$. The above distribution is of the form of
(\ref{exp}), with $z(y_t)=y_t^{\nu_t}$, $a(\phi_t)=\phi_t=1$,
$\gamma_t=-1/\lambda_t$, $b(\gamma_t)=-\log(-\nu_t\gamma_t)$ and
$c(y_t,\phi_t)=y_t^{\nu_t-1}$.

Given $\lambda_t$, the expectation and variance of $y_t^{\nu_t}$
are
$$
\E(y_t^{\nu_t}|\lambda_t)=\frac{\,db(\gamma_t)}{\,d\gamma_t}=\lambda_t
$$
and
$$
\var(y_t^{\nu_t}|\lambda_t)=\frac{\,d^2b(\gamma_t)}{\,d\gamma_t^2}=\lambda_t^2.
$$
Since $\lambda_t=\mu_t>0$, the logarithmic link
$g(\lambda_t)=\log\lambda_t=\eta_t$ can be used.

The prior and posterior distributions of $\lambda_t$ are inverted
gamma, i.e. $\lambda_t|y^{t-1}\sim IG(s_t-1,r_t)$ and
$\lambda_t|y^t\sim IG(s_t,r_t+y_t^{\nu_t})$ so that
$1/\lambda_t|y^{t-1}\sim G(s_t-1,r_t)$ and $1/\lambda_t|y^t\sim
G(s_t,r_t+y_t^{\nu_t})$, e.g.
$$
p(\lambda_t|y^{t-1})=\frac{r_t^{s_t-1}}{\Gamma(s_t-1)}
\frac{1}{\lambda_t^{s_t}} \exp\left(-\frac{r_t}{\lambda_t}\right).
$$
Since the link is logarithmic and the prior/posterior distributions
are inverted gamma, by writing $\log\lambda_t=-\log\lambda_t^{-1}$,
the approximation of $r_t$ and $s_t$ follow from a similar way as in
the Poisson, i.e.
$$
r_t=\frac{\exp(f_t)}{q_t} \quad \textrm{and} \quad
s_t=\frac{1+q_t}{q_t}
$$
and the posterior moments of $\log\lambda_t$ are given by
$$
f_t^*=\psi(s_t+y_t^{\nu_t}-1)-\log(r_t+1) \quad \textrm{and} \quad
q_t^*=\left.\frac{\,d\psi(x)}{\,dx}\right|_{x=s_t+y_t^{\nu_t}-1},
$$
which can be approximated by
$$
f_t^*\approx \log \frac{s_t+y_t^{\nu_t}-1}{r_t+1} +
\frac{1}{2(s_t+y_t^{\nu_t}-1)} \quad \textrm{and} \quad q_t^*
\approx \frac{2s_t+2y_t^{\nu_t}-3}{2(s_t+y_t^{\nu_t}-1)}.
$$

With the usual definition of $r_t(\ell)$ and $s_t(\ell)$ and their
calculation via $f_t(\ell)$, $q_t(\ell)$ and the above equation,
we obtain the $\ell$-step forecast distribution of $y_{t+\ell}$ as
\begin{equation}\label{weibull:for}
p(y_{t+\ell}|y^t)=\frac{r_t(\ell)^{s_t(\ell)-1}
y_{t+\ell}^{\nu_{t+\ell}-1} (s_t(\ell)-1) } {
(r_t(\ell)+y_{t+\ell}^{\nu_{t+\ell}})^{s_t(\ell)} }.
\end{equation}
Using conditional expectations, we can obtain the forecast mean
and variance of $y_{t+\ell}^{\nu_{t+\ell}}$ as
$$
y_t^{\nu_t}(\ell)=\E(y_{t+\ell}^{\nu_{t+\ell}}|y^t)=
\E(\E(y_{t+\ell}^{\nu_{t+\ell}}|\lambda_{t+\ell})|y^t)=\frac{r_t(\ell)}{s_t(\ell)-2},
$$
for $s_t(\ell)>2$ and
$$
\var(y_{t+\ell}^{\nu_{t+\ell}}|y^t) =
\E(\var(y_{t+\ell}^{\nu_{t+\ell}}|\lambda_{t+\ell})|y^t) +
\var(\E(y_{t+\ell}^{\nu_{t+\ell}}|\lambda_{t+\ell})|y^t) = \frac{
r_t(\ell)^2 (s_t(\ell)-1) }{ (s_t(\ell)-2)^2(s_t(\ell)-3) },
$$
for $s_t(\ell)>3$.

Considering a random walk evolution for
$\eta_t=\theta_t=\theta_{t-1}+\omega_t$, from the logarithmic
link, we obtain
\begin{equation}\label{weibull:lambda}
\lambda_t=\exp(\omega_t) \lambda_{t-1}
\end{equation}
and so $\lambda_t|\lambda_{t-1}\sim
LogN(\log\lambda_{t-1},\Omega)$, where $\Omega$ is the variance of
$\Omega$. The derivation of this result is the same as in the
Poisson example.

From (\ref{logl}) and $\lambda_t|\lambda_{t-1}\sim
LogN(\log\lambda_{t-1},\Omega)$, the log-likelihood function of
$\lambda_1,\ldots,\lambda_T$, based on data $y^T=\{y_1,\ldots,y_T\}$
is
$$
\ell(\lambda_1,\ldots,\lambda_T;y^T) = -\sum_{t=1}^T \left(
\frac{y_t^{\nu_t}}{\lambda_t}+\log\frac{\lambda_t}{\nu_t}+(1-\nu_t)
\log y_t +\frac{\log(2\pi\Omega)}{2}+
\frac{(\log\lambda_t-\log\lambda_{t-1})^2}{2\Omega}\right) .
$$

Power discounting yields
$$
r_{t+1}=\delta (r_t+y_t^{\nu_t}) \quad \textrm{and} \quad
s_{t+1}=\delta (s_t+1).
$$

We consider model comparison for the Weibull distribution when
$\eta_t=F\theta_t$ and $\theta_t=\theta_{t-1}+\omega_t$, for some
scalar $F$. This is an autoregressive type evolution for $\eta_t$.
We specify the variance of $\omega_t$ with a discount factor (West
and Harrison, 1997, Chapter 6) as
$\var(\omega_t)=\Omega_t=(1-\delta)P_{t-1}/\delta$, where $P_t$ is
the posterior variance of $\theta_t|y^t$. The density of
$y_t|y^{t-1}$ is given by (\ref{weibull:for}), for $\ell=1$,
$r_{t-1}(1)=r_t=\exp(f_t)/q_t$ and $s_{t-1}(1)=s_t=(1+q_t)/q_t$,
where $f_t=Fm_{t-1}$, $q_t=F^2P_{t-1}/\delta$ and $m_t$, $P_t$ are
updated from (\ref{post:th1}) as
$$
m_{t}=\log \frac{s_{t}+y_t-1}{r_{t}+1}+\frac{1}{2(s_{t}+y_t-1)}
$$
and
$$
P_t= \frac{P_{t-1}}{\delta}-\frac{P_{t-1}^2}{\delta^2} \left( 1-
\frac{ 2s_t+2y_t-3}{2(s_t+y_t-1)q_t}\right) \frac{1}{q_t} =
\frac{2s_t+2y_t-3}{2(s_t+y_t-1)F^2}.
$$

We consider now the situation of the choice of $\delta$. Suppose we
have two models $\mathcal{M}_1$ with a discount factor $\delta_1$
and $\mathcal{M}_2$ with $\delta_2$ and otherwise the models are the
same. The Bayes factor from a single observation ($k=1$) is given by
$$
H_t(1)= \frac{ r_{1t}^{s_{1t}-1} (s_{1t}-1)
(r_{2t}+y_t^{\nu_t})^{s_{2t}} }{ r_{2t}^{s_{2t}-1} (s_{2t}-1)
(r_{1t}+y_t^{\nu_t})^{s_{1t}}},
$$
where $r_{jt}$ and $s_{jt}$ are defined as $r_t$ and $s_t$ if we
replace $\delta$ by $\delta_j$, for $j=1,2$.

\begin{figure}[t]
 \epsfig{file=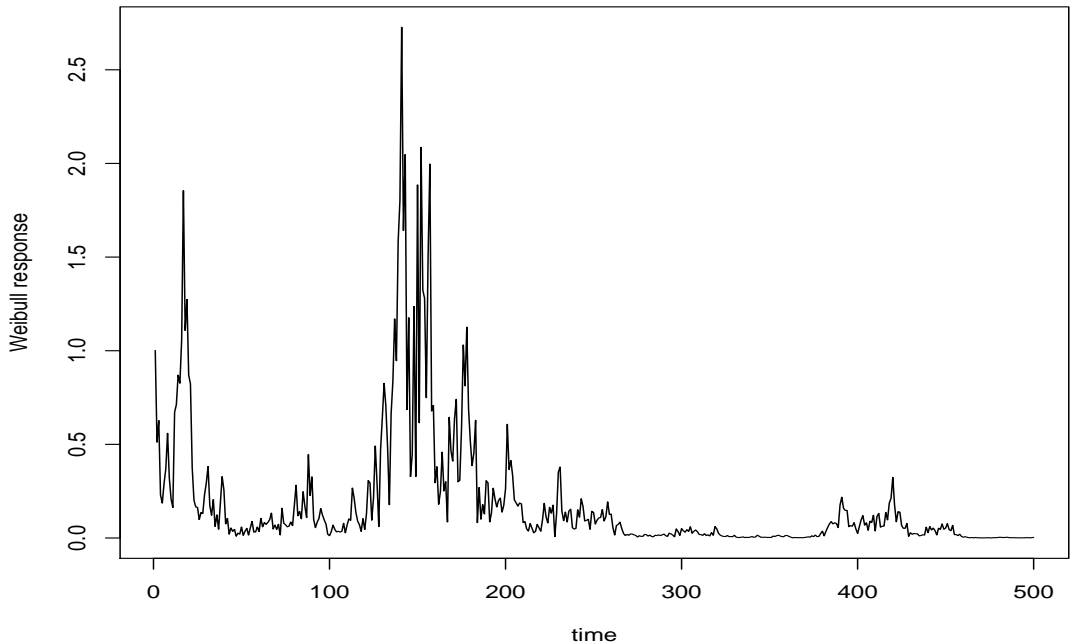, height=10cm, width=15cm}
 \caption{Simulated data from a Weibull distribution with $\nu_t=3$ and
 $\lambda_t$ generated from (\ref{weibull:lambda}).}\label{fig5}
\end{figure}

\begin{figure}[t]
 \epsfig{file=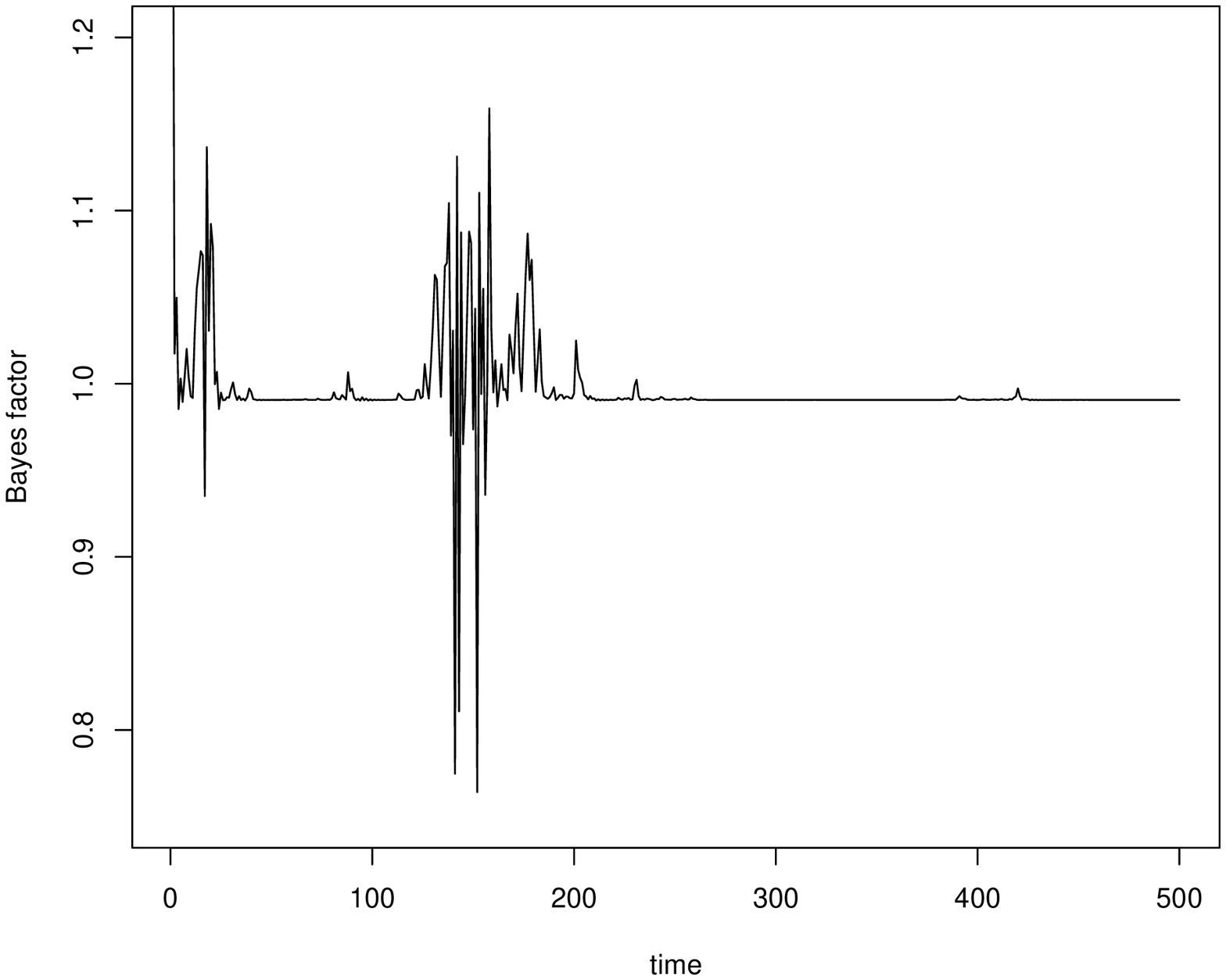, height=10cm, width=15cm}
 \caption{Bayes factor $\{H_t(1)\}$ of model $\mathcal{M}_1$ with $\delta=0.9$
 vs model $\mathcal{M}_2$ with $\delta_2=0.7$.}\label{fig6}
\end{figure}

For illustration, we simulate 500 observations from a Weibull
distribution with $\nu_t=3$ and $\{\lambda_t\}$ being simulated from
(\ref{weibull:lambda}), where we have used $F=1$, $\lambda_0=1$ and
$\omega_t\sim N(0,1)$. Figure \ref{fig5} shows the simulated data.
In order to choose the discount factor $\delta$, we apply the Bayes
factor $H_t(1)$ over a range values of $\delta_1,\delta_2\geq 0.5$.
We have used $m_0=0$ and a weakly informative prior $P_0=1000$.
Table \ref{table2} reports on $\bar{H}(1)$, the mean of $H_t(1)$,
and on the log-likelihood function
$\ell(\lambda_1,\ldots,\lambda_{500}|y^{500})$ evaluated at
$\widehat{\lambda}_t=(r_t+y_t^{\nu_t})/s_t$ (see the posterior
distribution of $\lambda_t|y^t$). This table indicates that there is
little difference in the performance of the one-step forecast
distribution, under the two models. The log-likelihood function
clearly indicates that $\delta_1=0.9$ produces the model with the
largest likelihood. The deficiency to separate the models using the
Bayes factor criterion, indicates that, in a sequential setting
which is appropriate for time series, one should better look at the
Bayes factor for each time $t$ and not at the overall mean of the
Bayes factor. Figure \ref{fig6} shows the Bayes factor of
$\mathcal{M}_1$ (with $\delta_1=0.9$) against $\mathcal{M}_2$ (with
$\delta_2=0.7$). We see that, although the mean of the Bayes factor
is 0.996 (see Table \ref{table2}), at $t=1-50$ and $t=100-200$,
there can be declared significant difference between the two models,
which is slightly in favour of model $\mathcal{M}_1$. This effect is
masked when one looks at the overall picture, considering the mean
$\bar{H}(1)$, and it indicates the benefit of sequential application
of Bayes factors.

\begin{table}
\caption{Log-likelihood function $\ell(.)$ and mean $\bar{H}(1)$ of
the Bayes factor sequence $\{H_t(1)\}$ of $\mathcal{M}_1$ (with
$\delta_1$) against $\mathcal{M}_2$ (with
$\delta_2$).}\label{table2}
\begin{center}
\begin{tabular}{|c||ccccccc|}
\hline & $\ell(.)$ & & & $\bar{H}(1)$ & & & \\
$\delta_1\backslash\delta_2$ & & 0.99 & 0.9 & 0.8 & 0.7 & 0.6 & 0.5
\\ \hline 0.99 & -5.787 & 1 & 0.997 & 0.995 & 0.994 & 0.994 & 0.998
\\ 0.95 & -7.411 & 1.001 & 0.999 & 0.997 & 0.995 & 0.995 &
0.999 \\ 0.90 & \textbf{-3.123} & 1.002 & 1 & 0.998 & 0.996 & 0.996 & 1 \\
0.85 & -8.547 & 1.004 & 1.001 & 0.999 & 0.997 & 0.997 & 1.001 \\
0.80 & -8.854 & 1.005 & 1.002 & 1 & 0.998 & 0.998 & 1.002 \\ 0.75 &
-9.098 & 1.006 & 1.003 & 1.001 & 0.999 & 0.999 & 1.002 \\ 0.70 &
-9.301 & 1.007 & 1.004 & 1.002 & 1 & 0.999 & 1.003 \\ 0.65 & -9.476
& 1.008 & 1.005 & 1.002 & 1 & 1 & 1.003 \\ 0.6 & -9.631 & 1.008 &
1.005 & 1.003 & 1.001 & 1 & 1.003 \\ 0.55 & -9.771 & 1.008 & 1.005 &
1.003 & 1 & 0.999 & 1.002 \\ 0.50 & -9.947 & 1.007 & 1.004 & 1.001 &
0.998 & 0.997 & 1 \\
\hline
\end{tabular}
\end{center}
\end{table}

The exponential and Weibull distributions are useful models for the
analysis of survival times data. In the context of DGLMs, we have
dynamic survival models due to Gamerman (1991). Here we give a brief
description of dynamic survival models and we extend a result of
Gamerman (1991).

Suppose that, given $\nu_t$ and $\lambda_t$, the survival time $y_t$
follows the Weibull distribution $p(y_t|\lambda_t)$ (here we assume
that $\nu_t$ is known and so we exclude it from conditioning). For
example, if the exponential distribution is believed to be an
appropriate model, we have $\nu_t=1$. The survivor function of the
Weibull distribution is
\begin{equation}\label{survival1}
S(y_t|\lambda_t)=\frac{\nu_t}{\lambda_t}\int_{y_t}^\infty
u_t^{\nu_t-1} \exp\left( - \frac{u_t^{\nu_t}}{\lambda_t}\right)\,d
u_t = \exp\left( -\frac{y_t^{\nu_t}}{\lambda_t}\right).
\end{equation}
Suppose we have a vector of $p$ regressor variables or covariates
$x=[x_1~\cdots~x_p]'$ and we consider a vector of parameters $\beta$
so that $1/\lambda_t$ is proportional to $\exp(x'\beta)$. Then the
hazard function $h(y_t;\nu_t,\lambda_t)\equiv h(t)\propto
\nu_ty_t^{\nu_t-1} \exp(x_t'\beta)$ and this leads to the
proportional hazards model with $h(t)=h_0(t) \exp(x'\beta)$, where
$h_0(t)$ is the baseline hazard function (Dobson, 2002, \S10.2). So
one can write $\log h(t)=\log h_0(t) +x'\beta$ and considering a
partition of $(0,N)$ as $0=y_0<y_1<\cdots<y_T=N$ so that $t\in
I_t=(y_{t-1},y_t]$, we write $\log h_0(t)=\alpha_t$, i.e. the
baseline is a step function that takes a constant value $\alpha_t$
at each time interval $I_t$.

Now in the DGLM flavor, dynamic survival models assume that $\beta$
evolves over time between intervals $I_1,\ldots,I_T$, but it remains
constant inside each interval $I_t$. Gamerman (1991) considers the
model
\begin{equation}\label{survival2}
\log \lambda_t^{(j)}=\log h^{(j)}(t) = F_j' \theta_t, \quad
j=1,\ldots,i_t; \quad t=1,\ldots,T,
\end{equation}
where $F_j=[1~x_j']'$ is the design vector and
$\theta_t=[\alpha_t~\beta_t']'$ is the time-varying parameter
vector, which is assumed to follow a random walk evolution according
to $\theta_t=\theta_{t-1}+\omega_t$, and $\lambda_t$ has been
modified to $\lambda_t^{(j)}$ to account for individual $j$. Here,
$t$ indexes the $T$ intervals $I_1,\ldots,I_T$ of $(0,N)$ and $j$
indexes each individual to be alive at the beginning of $I_t$, where
$i_t$ is the number of such individuals in $I_t$. Note that through
$x_j$, each individual $j$ may have different effects through
different regressor variables, although it is not unrealistic to set
$x_j=x$ or $F=[1~x']'$ (for all individuals we have the same
regressor variables). The dynamics of the system is reflected on the
dynamics of $\theta_t$. Equations (\ref{survival1}) and
(\ref{survival2}) define a dynamic survival model, which Bayesian
inference follows, in an obvious extension of the DGLM estimation,
providing the posterior first two moments of $h^{(j)}(t)$ (details
appear in Gamerman, 1991).

Fix individual $j$ and write $\lambda_t^{(j)}=\lambda_t$. Given the
adopted random walk evolution for $\theta_t$, for any $y_t^*\in
I_t=(y_{t-1},y_t]$, the prior $\lambda_t^{-1}|y^{t-1}\sim
G(s_t-1,r_t)$ combines with the survivor function (\ref{survival1})
to give the survivor prediction
\begin{eqnarray*}
S(y_t^*|y^{t-1}) &=& \int_0^\infty
S((y_t^*-y_{t-1})|\lambda_t)p(\lambda_t^{-1}|y^{t-1})\,d\lambda_t^{-1}
\\ &=& \frac{r_t^{s_t-1}}{\Gamma(s_t-1)} \int_0^\infty
\lambda_t ^{-(s_t-1)} \exp ( -
((y_t^*-y_{t-1})^{\nu_t}+r_t)\lambda_t^{-1} ) \,d\lambda_t^{-1} \\
&=& \left( 1+ \frac{(y_t^*-y_{t-1})^{\nu_t}}{r_t} \right)
^{-(s_t-1)},
\end{eqnarray*}
where we can see that for $\nu_t=1$, we obtain the survivor
prediction of the exponential distribution, reported in Gamerman
(1991). Thus $S(y_t^*|y^{t-1})$ predicts the remaining survival time
of individual $j$ still alive.

\subsubsection{Pareto and beta}

The Pareto (Johnson {\it et al.}, 1994) is a skewed distribution
with many applications in social, scientific and geophysical
phenomena. For example, in economics it can describe the allocation
of wealth among individuals or prices of the returns of stocks.

Suppose that the time series $\{y_t\}$ is generated from Pareto
distribution with density
$$
p(y_t|\lambda_t)=\lambda_ty_t^{-\lambda_t-1}, \quad y_t\geq 1;
\quad \lambda_t>0.
$$
This distribution is also known as Pareto(I) distribution and
$\lambda_t$ is known as the index of inequality (this distribution
is examined in detail in Johnson {\it et al.}, 1994). The above
distribution is of the form of (\ref{exp}), with $z(y_t)=\log y_t$,
$a(\phi_t)=\phi_t=1$, $\gamma_t=-\lambda_t$,
$b(\gamma_t)=-\log(-\gamma_t)$ and $c(y_t,\phi_t)=1/y_t$. We note
that by setting $x_t=1/y_t$ or $x_t=1/(1-y_t)$, we have that
$0<x_t<1$ so that, given $\lambda_t$, $x_t$ follows a beta
distribution with parameters $\lambda_t,1$ and $1,\lambda_t$,
respectively. Thus inference for the Pareto distribution can be
readily applied to the beta distribution (Johnson {\it et al.},
1994) when at least one parameter of the beta distribution is equal
to 1. This is a useful consideration as we can deal with responses
being proportions or probabilities.

We have
$$
\E(y_t|\lambda_t)=\frac{\lambda_t}{\lambda_t-1}=\mu_t \quad
(\lambda_t>1) \quad \textrm{and} \quad
\var(y_t|\lambda_t)=\frac{\lambda_t}{(\lambda_t-1)^2 (\lambda_t-2)}
\quad (\lambda_t>2).
$$

Since $\mu_t>0$, the logarithmic link function can be used, so that
$g(\mu_t)=\log\mu_t=\log\lambda_t-\log(\lambda_t-1)$, for
$\lambda_t>1$. Using the transformation $\gamma_t=-\lambda_t$, we
find that the prior and posterior distributions of $\lambda_t$ are
gamma, i.e. $\lambda_t|y^{t-1}\sim G(s_t+1,r_t)$ and
$\lambda_t|y^t\sim G(s_t+2,r_t+\log y_t)$, respectively.

Following the approximation of $r_t$ and $s_t$ in the Poisson case,
we have that
$$
r_t=\frac{\exp(-f_t)}{q_t} \quad \textrm{and} \quad
s_t=\frac{1-q_t}{q_t}
$$
and the posterior moments of $\log\lambda_t$ are given by
$$
f_t^*=\psi(s_t+\log y_t+1) -\log(r_t+1) \quad \textrm{and} \quad
q_t^*=\left.\frac{\,d\psi(x)}{\,dx}\right|_{x=s_t+\log y_t +1},
$$
which can be approximated by
$$
f_t^*\approx \log \frac{s_t+\log y_t+1}{r_t+1} + \frac{1}{2(s_t+\log
y_t+1)} \quad \textrm{and} \quad q_t^*=\frac{2s_t+2\log
y_t+1}{2(s_t+\log y_t+1)}.
$$

Power discounting yields
$$
r_{t+1}=\delta(r_t+\log y_t) \quad \textrm{and} \quad
s_{t+1}=\delta(s_t+1).
$$

With $r_t(\ell)$ and $s_t(\ell)$ computed from $f_t(\ell)$ and
$q_t(\ell)$ and the above equations of $r_t$ and $s_t$, the
$\ell$-step forecast distribution of $y_{t+\ell}$ is
$$
p(y_{t+\ell}|y^t)=\frac{r_t(\ell)^{s_t(\ell)+1} (s_t(\ell)+1)}{
y_{t+\ell}(r_t(\ell)+\log y_{t+\ell})^{s_t(\ell)+1} }.
$$

Considering a random walk evolution for
$\eta_t=\theta_t=\theta_{t-1}+\omega_t$, we have that the evolution
of $\lambda_t$ is
$$
\lambda_t=\frac{\lambda_{t-1}\exp(\omega_t)}{\lambda_{t-1}\exp(\omega_t)-\lambda_{t-1}+1},
$$
from which we can obtain the distribution of
$\lambda_t|\lambda_{t-1}$. With this, assuming that $\omega_t\sim
N(0,\Omega)$ and that $\lambda_t>1$, the density of
$\lambda_t|\lambda_{t-1}$ is
$$
p(\lambda_t|\lambda_{t-1})=\frac{1}{\sqrt{2\pi\Omega}
\lambda_t(\lambda_t-1)} \exp\left( -\frac{1}{2\Omega} \left( \log
\frac{\lambda_t(\lambda_{t-1}-1)}{\lambda_{t-1}(\lambda_t-1)}\right)^2\right),
$$
where $\Omega$ should be chosen so that to guarantee $\lambda_t>1$,
for all $t$. Then from (\ref{logl}) the log-likelihood function is
\begin{eqnarray*}
\ell(\lambda_1,\ldots,\lambda_T;y^T) &=& \sum_{t=1}^T \bigg(
-\lambda_t\log y_t +\log\lambda_t-\log y_t \\ &&
-\log\sqrt{2\pi\Omega}\lambda_t(\lambda_t-1) - \frac{1}{2\Omega}
\left(
\log\frac{\lambda_t(\lambda_{t-1}-1)}{\lambda_{t-1}(\lambda_t-1)}\right)^2
\bigg),
\end{eqnarray*}
for $\lambda_1,\ldots,\lambda_T>1$.

Bayes factors can be computed from the predictive density
$p(y_{t+1}|y^t)$ and (\ref{bf1}). As an example consider the
comparison of two models $\mathcal{M}_1$ and $\mathcal{M}_2$, which
differ in some quantitative aspects, e.g. in the discount factor
$\delta$ (see also the illustration that follows). By defining
$r_{jt}$ and $s_{jt}$ the respective values of $r_t$ and $s_t$, for
model $\mathcal{M}_j$ $(j=1,2)$, the Bayes factor $H_t(1)$ can be
expressed as
$$
H_t(1)= \frac{ r_{1,t+1}^{s_{1,t+1}+1} (s_{1,t+1}+1) (r_{2,t+1}+\log
y_{t+1})^{s_{1,t+1}+1} } { r_{2,t+1}^{s_{2,t+1}+1} (s_{2,t+1}+1)
(r_{1,t+1}+\log y_{t+1})^{s_{2,t+1}+1} }.
$$

\begin{figure}[t]
 \epsfig{file=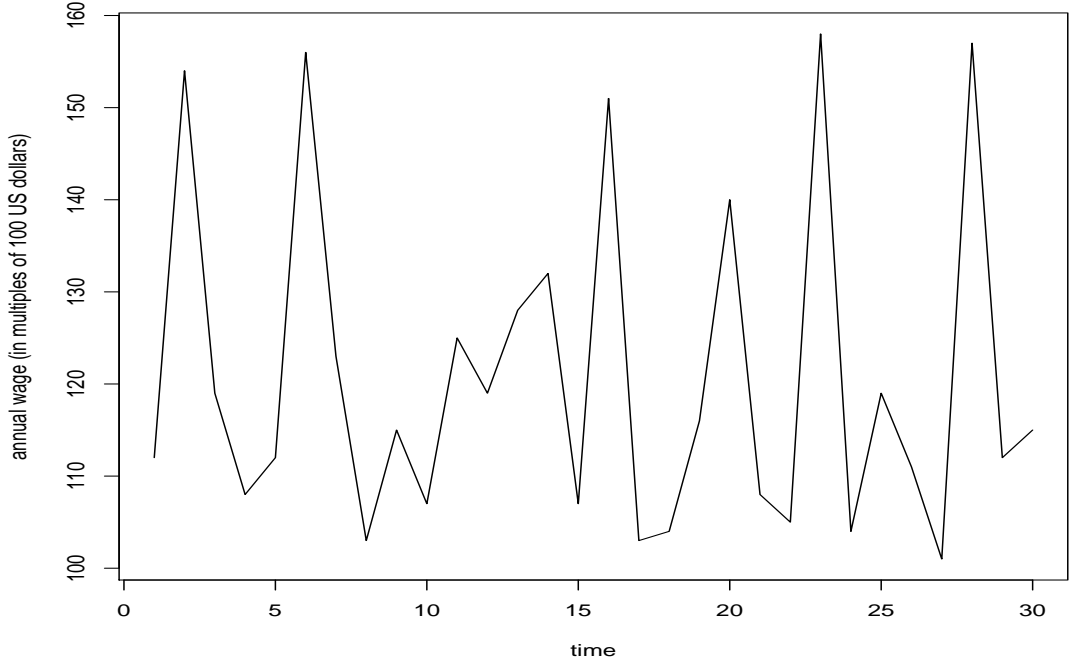, height=10cm, width=15cm}
 \caption{Annual wage Pareto data.}\label{fig:pareto1}
\end{figure}

\begin{figure}[h]
 \epsfig{file=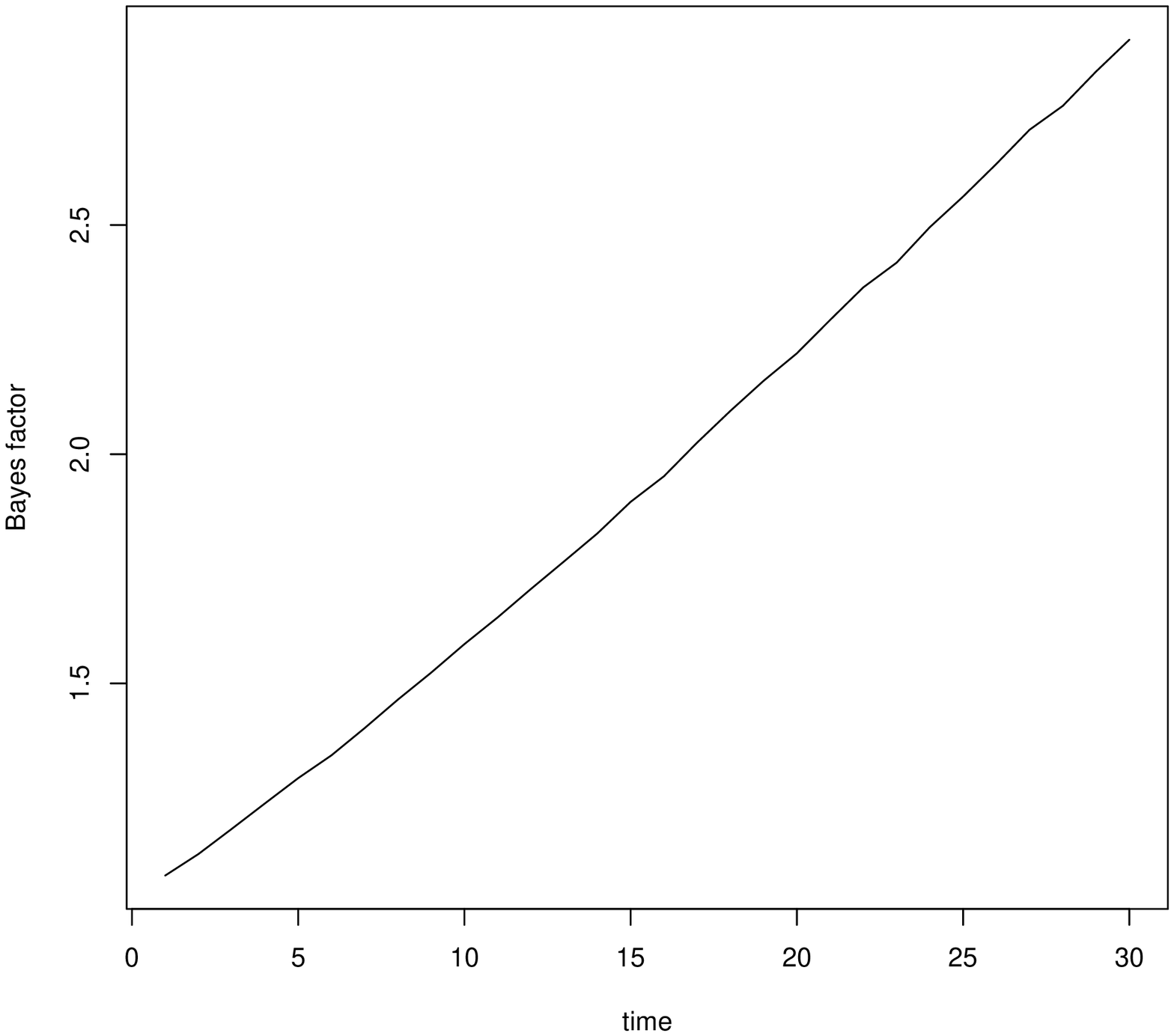, height=10cm, width=15cm}
 \caption{Bayes factor $\{H_t(1)\}$ of model $\mathcal{M}_1$ with $\delta=0.99$
 vs model $\mathcal{M}_2$ with $\delta_2=0.95$ for the Pareto data.}\label{fig:pareto2}
\end{figure}

To illustrate the above Pareto model for time series data, we
consider the data of Arnold and Press (1989), consisting of 30 wage
observations (in multiples of US dollars) of production-line workers
in a large industrial firm; the data are also discussed in Dyer
(1981). The data are shown in Figure \ref{fig:pareto1}, from which
two points can be argued it: (a) the data appear to be
autocorrelated (in fact it is easy to run a corrolagram to justify
this) and (b) the data exhibit a local level behaviour (one could
argue for local stationarity, but with only 30 observations a local
level model seems more appropriate). Here we apply the Pareto model
with $r_t$ and $s_t$ being updated by the power discounting (this is
appropriate for the local level behaviour of the time series). Table
\ref{table3} shows the mean of the Bayes factors for various values
of the discount factors $\delta_1$ and $\delta_2$ in the range of
$[0.5,0.99]$. It is evident that the best model is the model with
$\delta=0.99$, which is capable of producing Bayes factors larger
than 1 as compared with models with lower discount factors. From
that table it is also evident that models with low discount factors
do worse than models with high discount factors and so by far the
worst model is that using $\delta=0.5$. Figure \ref{fig:pareto2}
shows the values of the Bayes factor of the model with $\delta=0.99$
against the model with $\delta=0.95$; we note that all values of the
Bayes factor are larger than one and there is a steady increase in
the Bayes factors indicating the superiority of the model with
$\delta=0.99$.

\begin{table}
\caption{Mean $\bar{H}(1)$ of the Bayes factor sequence $\{H_t(1)\}$
of $\mathcal{M}_1$ (with $\delta_1$) against $\mathcal{M}_2$ (with
$\delta_2$) for the Pareto model.}\label{table3}
\begin{center}
\begin{tabular}{|c||cccccc|}
\hline & & & $\bar{H}(1)$ & & & \\
$\delta_1\backslash\delta_2$ & 0.99 & 0.9 & 0.8 & 0.7 & 0.6 & 0.5
\\ \hline 0.99 & 1 & 1.950 & 3.484 & 5.414 & 7.786 & 10.798 \\ 0.95
& 0.749 & 1.401 & 2.449 & 3.774 & 5.409 & 7.489 \\ 0.90 & 0.559 & 1
& 1.708 & 2.608 & 3.721 & 5.141 \\ 0.85 & 0.439 & 0.760 & 1.276 &
1.931 & 2.745 & 3.785 \\ 0.80 & 0.358 & 0.605 & 1 & 1.503 & 2.129 &
2.931 \\ 0.75 & 0.299 & 0.496 & 0.810 & 1.211 & 1.711 & 2.350 \\
0.70 & 0.254 & 0.415 & 0.672 & 1 & 1.408 & 1.932 \\ 0.65 & 0.218 &
0.352 & 0.566 & 0.839 & 1.179 & 1.616 \\ 0.60 & 0.189 & 0.302 &
0.482 & 0.712 & 1 & 1.368 \\ 0.55 & 0.164 & 0.261 & 0.414 & 0.609 &
0.854 & 1.167 \\ 0.50 & 0.143 & 0.225 & 0.356 & 0.523 & 0.732 & 1\\
\hline
\end{tabular}
\end{center}
\end{table}

\subsubsection{Inverse Gaussian}

The inverse Gaussian or Wald (Chhikara and Folks, 1989; Johnson {\it
et al.}, 1994) is a skewed distribution that can describe phenomena
in economics and in many other sciences. This distribution is known
as the first passage time distribution of Brownian motion with
positive drift. Recently, Huberman {\it et al.} (1998) used an
inverse Gaussian distribution to model internet flow and internet
traffic.

Suppose that the time series $\{y_t\}$ is generated from an inverse
Gaussian distribution, that is for given $\mu_t$ and $\lambda_t$,
the density function of $y_t$ is
$$
p(y_t|\mu_t,\lambda_t) = \sqrt{ \frac{\lambda_t}{2\pi y_t^3} } \exp
\left( - \frac{\lambda_t (y_t-\mu_t)^2 }{ 2\mu_t^2 y_t} \right),
\quad y_t>0; \quad \mu_t,\lambda_t>0.
$$
This is a unimodal distribution, which converges to the normal
distribution, as $\lambda_t\rightarrow\infty$. To the following we
will assume that $\lambda_t$ is a known parameter and interest will
be placed on $\mu_t$; hence we write $p(y_t|\mu_t,\lambda_t)\equiv
p(y_t|\mu_t)$. We can see that the above distribution is of the form
of (\ref{exp}), with $z(y_t)=y_t$, $\phi_t=\lambda_t$,
$a(\phi_t)=2/\lambda_t$, $\gamma_t=-1/\mu_t^2$,
$b(\gamma_t)=-2/\mu_t=-2\sqrt{-\gamma_t}$ and
$c(y_t,\phi_t)=(\lambda_t/(2\pi y_t^3))^{1/2}
\exp(-\lambda_t/(2y_t))$. Then we can verify that
$$
\E(y_t|\mu_t)=\frac{\,db(\gamma_t)}{\,d\gamma_t}=\frac{1}{\sqrt{-\gamma_t}}=\mu_t
$$
and
$$
\var(y_t|\mu_t)=a(\phi_t) \frac{\,d^2b(\gamma_t)}{\,d\gamma_t^2} =
\frac{a(\phi_t)}{2\sqrt{-\gamma_t^3}}=\frac{\mu_t^3}{\lambda_t}.
$$

The canonical link maps $\mu_t$ to $\gamma_t$, or
$g(\mu_t)=\gamma_t=-1/\mu_t^2$, but this is not convenient, since
$g(\mu_t)<0$ and hence we need to find an appropriate definition of
$F$ and $G$ in the state space representation of $g(\mu_t)=\eta_t$
in order to guarantee $-\infty<\eta_t<\infty$. The logarithmic link,
$g(\mu_t)=\log\mu_t$, seems to work better, since it maps $\mu_t$ to
the real line and so $F'\theta_t=\eta_t=g(\mu_t)$ is defined easily.

The prior distribution of $\mu_t$ can be defined via the prior
distribution of $\gamma_t$ and the transformation
$\gamma_t=-1/\mu_t^2$. In the appendix it is shown that
\begin{equation}\label{eq:igaussian:2}
p(\mu_t|y^{t-1})= \frac{ 2\exp(s_t^2/r_t) r_t }{ (\exp(s_t^2/r_t)
s_t \sqrt{\pi/r_t} + 1)\mu_t^3}  \exp\left( -
\frac{(r_t-\mu_ts_t)^2}{r_t\mu_t^2}\right).
\end{equation}
In the appendix it is shown that
\begin{equation}\label{IG:prior:m}
\E(\mu_t|y^{t-1}) = \frac{ \sqrt{\pi r_t} \exp(s_t^2/r_t) }{
\exp(s_t^2/r_t)s_t\sqrt{\pi/r_t}+1}.
\end{equation}

The posterior distribution of $\mu_t$ is obtained from the posterior
distribution of $\gamma_t$ as
\begin{eqnarray*}
p(\mu_t|y^t) &=& \kappa(r_t+\lambda_ty_t, s_t+\lambda_t) \exp\left(
-\frac{r_t+\lambda_ty_t}{\mu_t^2} +
\frac{2(s_t+\lambda_t)}{\mu_t}\right) \frac{2}{\mu_t^3} \\ &=&
\frac{ 2 \exp ( (s_t+\lambda_t)^2/(r_t+\lambda_ty_t))
(r_t+\lambda_ty_t) }{ ( \exp ( (s_t+\lambda_t)^2/(r_t+\lambda_ty_t))
(s_t+\lambda_t) \sqrt{\pi / (r_t+\lambda_ty_t)} + 1 ) \mu_t^3} \\ &&
\times \exp \left( -
\frac{(r_t+\lambda_ty_t-\mu_t(s_t+\lambda_t))^2}{ (r_t+\lambda_t
y_t)\mu_t^2}\right),
\end{eqnarray*}
where in the appendix it is shown that
$$
\kappa(r_t,s_t)=r_t\left( \exp\left(\frac{s_t^2}{r_t}\right) s_t
\sqrt{\frac{\pi}{r_t}} + 1\right)^{-1}.
$$
The approximation of $r_t$ and $s_t$ is difficult, since the moment
generating function of $\eta_t=\log\mu_t$ (which is needed in order
to compute $r_t$ and $s_t$) is not available in close form. Thus
power discounting should be applied. From the posterior of
$\gamma_t|y^t$, given by (\ref{post:g1}), we have
$$
(p(\gamma_t|y^t))^\delta \propto \exp\left( \delta \left( r_t+
\frac{2y_t}{\lambda_t}\right)\gamma_t+2\delta \left(
s_t+\frac{2}{\lambda_t}\right)\sqrt{-\gamma_t}\right)
$$
and so from the prior of $\gamma_{t+1}$ (equation (\ref{prior:g1}))
and the power discounting law we obtain
$$
r_{t+1}=\frac{\delta (r_t\lambda_t +2y_t)}{\lambda_t} \quad
\textrm{and} \quad s_{t+1}= \frac{\delta
(s_t\lambda_t+2)}{\lambda_t}.
$$
With $r_t(\ell)=r_{t+1}$ and $s_t(\ell)=s_{t+1}$, the $\ell$-step
forecast distribution of $y_{t+\ell}|y^t$ is
\begin{eqnarray*}
p(y_{t+\ell}|y^t) &=& c (r_{t+1}+2y_{t+\ell})^{-1}
\frac{1}{\sqrt{y_{t+\ell}^3}} \exp
\left(-\frac{\lambda_{t+\ell}}{2y_{t+\ell}}\right) \left(
\frac{s_{t+1}\lambda_{t+\ell}+2}{\lambda_{t+\ell}} \right. \\ &&
\left.\times \exp \left( \frac{ (s_{t+1}\lambda_{t+\ell} + 2)^2}{
\lambda_{t+\ell} ( r_{t+1} \lambda_{t+\ell}+2y_{t+\ell})} \right)
 \sqrt{
\frac{\lambda_{t+\ell}\pi}{r_{t+1}\lambda_{t+\ell}+2y_{t+\ell}}} + 2
\right),
\end{eqnarray*}
where the normalizing constant $c$ is
$$
c=(2\pi)^{-1/2}\sqrt{\lambda_{t+\ell}^3}r_{t+1} \left( s_{t+1}
\exp\left(\frac{s_{t+1}^2}{r_{t+1}}\right)\sqrt{\frac{\pi}{r_{t+1}}}+1\right)^{-1}.
$$
The $\ell$-step forecast mean can be deduced by (\ref{IG:prior:m})
as
$$
\E(y_{t+\ell}|y^t)=\E(\E(y_{t+\ell}|\mu_{t+\ell})|y^t)=\E(\mu_{t+\ell}|y^t)
= \frac{ \sqrt{\pi r_t(\ell)} \exp(s_t(\ell)^2/r_t(\ell))}{
\exp(s_t(\ell)^2/r_t(\ell))s_t(\ell)\sqrt{\pi / r_t(\ell)}+1}
$$

Of course the above power discounting specifies $r_t$ and $s_t$, for
a random walk type evolution for the prior (\ref{eq:igaussian:2}).
Following this, we can specify
$\log\mu_t=\eta_t=\theta_t=\theta_{t-1}+\omega_t$, with
$\omega_t\sim N(0,\Omega)$, and so
$$
\mu_t=\mu_{t-1}\exp(\omega_t),
$$
which leads to the density
$$
p(\mu_t|\mu_{t-1})=\frac{1}{\sqrt{2\pi\Omega}\mu_t} \exp\left( -
\frac{(\log\mu_t-\log\mu_{t-1})^2}{2\Omega}\right).
$$
Therefore, using (\ref{logl}), the log-likelihood function is
\begin{eqnarray*}
\ell(\mu_1,\ldots,\mu_T;y^T) &=& \sum_{t=1}^T \bigg(
\frac{\lambda_t}{2\mu_t^2}(2\mu_t-y_t)
+\log\sqrt{\frac{\lambda_t}{2\pi y_t^3}} - \frac{\lambda_t}{2y_t} \\
&& -\log\sqrt{2\pi\Omega}\mu_t -
\frac{(\log\mu_t-\log\mu_{t-1})^2}{2\Omega} \bigg).
\end{eqnarray*}
Bayes factors can be easily computed from $p(y_{t+1}|y^t)$ and the
Bayes factor formula (\ref{bf1}).

\begin{figure}[t]
 \epsfig{file=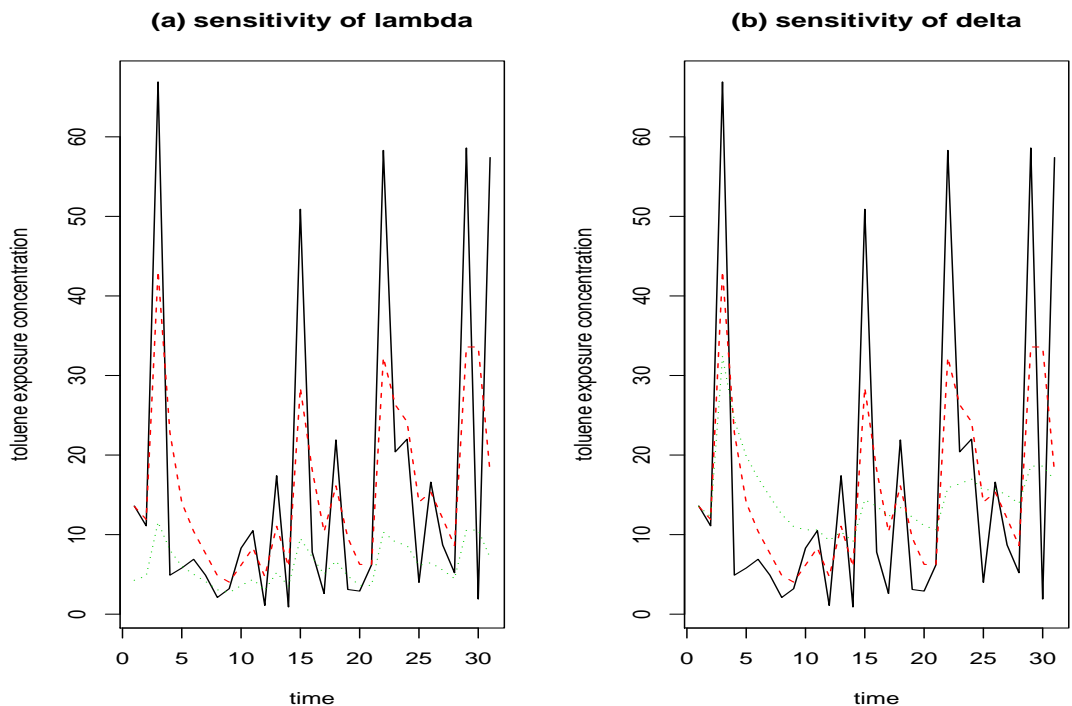, height=10cm, width=15cm}
 \caption{One-step forecast mean for the TEC data; panel (a) shows the
 actual data (solid line), the one-step forecasts with $\delta=0.5$ and $\lambda=0.01$ (dashed line),
 and the one-step forecasts with $\delta=0.5$ and $\lambda=1$ (dotted line); panel (b) shows
 the actual data (solid line), the one-step forecasts with $\delta=0.5$ and $\lambda=0.01$ (dashed line),
 and the one-step forecasts with $\delta=0.9$ and $\lambda=0.01$ (dotted line).}\label{fig:ig1}
\end{figure}

\begin{figure}[h]
 \epsfig{file=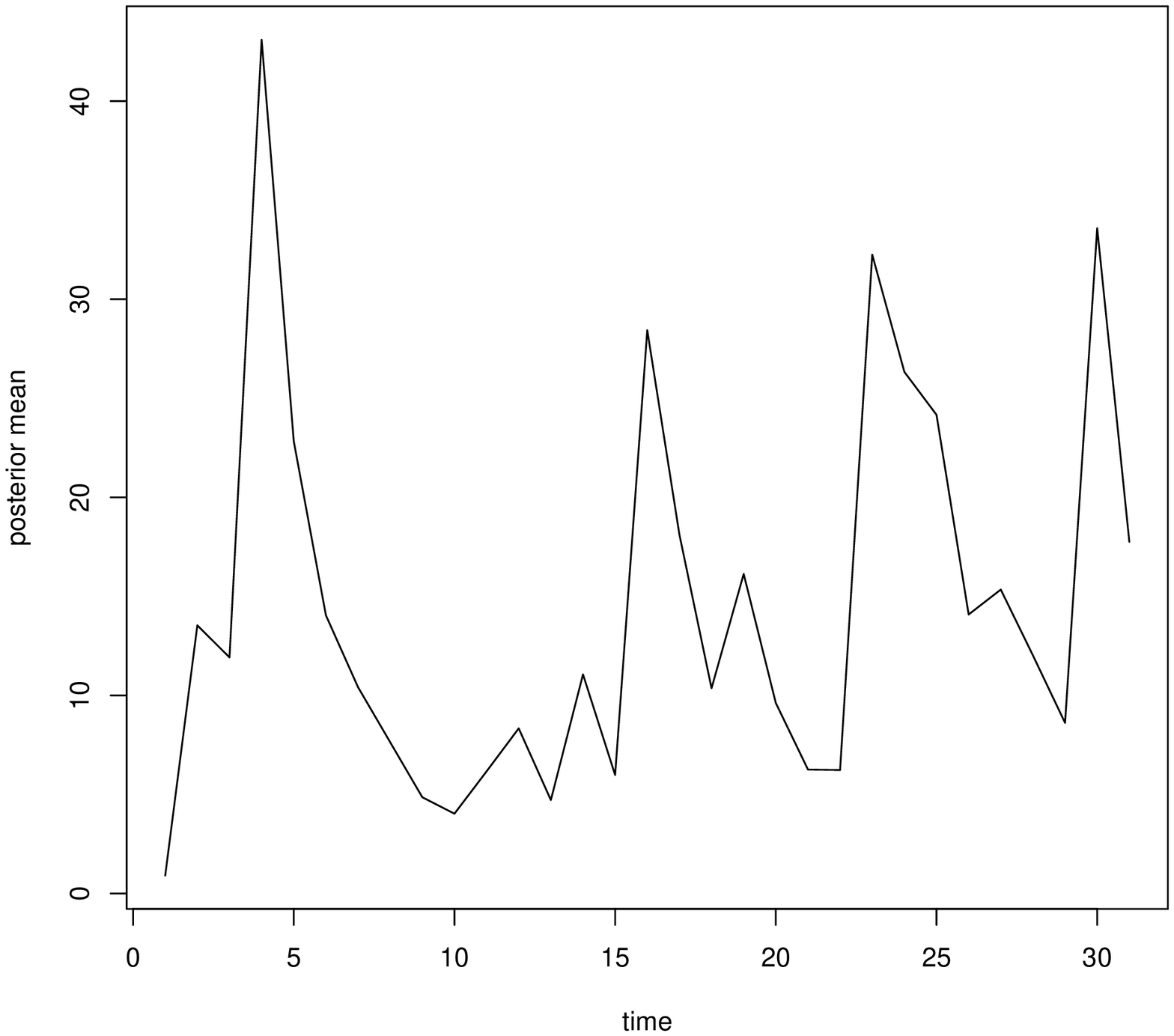, height=10cm, width=15cm}
 \caption{Posterior mean $\{\E(\mu_t|y^t)\}$ of the ETC data.}\label{fig:ig2}
\end{figure}

To illustrate the inverse Gaussian distribution we consider data
consisting of 30 daily observations of toluene exposure
concentrations (TEC) for a single worker doing stain removing. The
data can be found in Takagi {\it et al.} (1997) who propose a simple
model fit using maximum likelihood estimation for the inverse
Gaussian distribution. However, it may be argued that these data are
autocorrelated and so an appropriate time series should be fitted.
Figure \ref{fig:ig1} shows one-step forecasts means against the TEC
data. The forecast means are computed using the above DGLM model for
the inverse Gaussian response, using $\lambda_t=\lambda$. The
results show that a low value of the discount factor $\delta=0.5$
and a low value of $\lambda=0.01$ yield the best forecasts. The
posterior mean $\E(\mu_t|y^t)$ is plotted in Figure \ref{fig:ig2},
from which we can clearly see that there is a time-varying feature
of the parameters of the inverse Gaussian distribution. This is
failed to be recognized in Takagi {\it et al.} (1997). These authors
propose estimates for the mean and the scale of the inverse Gaussian
distribution as 16.7 and 6.4, which are both larger than the mean of
the posterior means
$(\E(\mu_1|y^1)+\cdots+\E(\mu_{30}|y^{30}))/30=14.48$ and
$\lambda=0.01$. We note that from Figure \ref{fig:ig1} as $\lambda$
increases, the forecast performance deteriorates so that a value of
$\lambda$ near 6.4 would yield poor forecast accuracy. The model we
propose here exploits the dynamic behaviour of $\mu_t$ and it is an
appropriate model for forecasting.

\section{Concluding comments}\label{discussion}

In this paper we discuss approximate Bayesian inference of dynamic
generalized linear models (DGLMs), following West {\it et al.}
(1985) and co-authors. Such an approach allows the derivation of the
multi-step forecast distribution, which is a useful consideration
for carrying out error analysis based on residuals, on the
likelihood function, or on Bayes factors. We explore all the above
issues by examining in detail several examples of distributions
including binomial, Poisson, negative binomial, geometric, normal,
log-normal, gamma, exponential, Weibull, Pareto, two special cases
of the beta, and inverse Gaussian.

We believe that DGLMs offer a unique statistical framework for
dealing with a range of statistical problems, including business and
finance, medicine, biology and genetics, and behavioural sciences.
In most of these areas, researchers are not well aware of the
advantages that Bayesian inference for DGLMs can offer. In this
context we believe that the present paper offers a clear description
of the methods with detailed examples of many useful response
distributions.

\renewcommand{\theequation}{A-\arabic{equation}} 
\setcounter{equation}{0}  
\appendix
\section*{Appendix}

\subsection*{Proof of equations (\ref{eq:binom:rt}) and
(\ref{eq:poisson:rt})}

First we calculate the mean and variance of the log-gamma and the
log-beta distributions. Let $X$ follow the gamma distribution with
parameters $\alpha$ and $\beta$, with density function
$$
p(x)=\frac{\beta^{\alpha}}{\Gamma(\alpha)}x^{\alpha-1}\exp(-\beta
x),
$$
where $\Gamma(.)$ denotes the gamma function and $\alpha,\beta>0$.
The density function of $Y=\log X$ is
$$
p(y)=\frac{\beta^{\alpha}}{\Gamma(\alpha)} \exp((\alpha
y)-\beta\exp(y)).
$$
The moment generating function of $Y$ is
$$
M_Y(z)=\E(\exp(zY))=\int_{-\infty}^{\infty}
\frac{\beta^{\alpha}}{\Gamma(\alpha)} \exp((\alpha+z)y-\beta\exp(y))
\,dy = \frac{\Gamma(\alpha+z)}{\Gamma(\alpha)\beta^z}
$$
and the cumulant generating function is $K_Y(z)=\log M_Y(z)=\log
\Gamma(\alpha+z)-\log\Gamma(\alpha)-z\log\beta$. Then we have
\begin{equation}\label{log:gamma1}
\E(Y)=\left. \frac{\,dK(z)}{\,dz}
\right|_{z=0}=\psi(\alpha)-\log\beta \quad \textrm{and} \quad
\var(Y)=\left. \frac{\,d^2K(z)}{\,dz^2} \right|_{z=0} =
\frac{\,d\psi(\alpha)}{\,d\alpha},
\end{equation}
where $\psi(.)$ is the digamma function, which is defined by
$\psi(x)=\,d\log\Gamma(x)/\,dx$ and the derivative $\psi(.)$ is
known as the trigamma function (Abramowitz and Stegun, 1964).

For the log-beta distribution, let $X$ follow the beta distribution,
with density function
$$
p(x)=\frac{\Gamma(\alpha+\beta)}{\Gamma(\alpha)\Gamma(\beta)}
x^{\alpha-1} (1-x)^{\beta-1},
$$
where $\alpha,\beta>0$ and $0<x<1$. The density function of $Y=\log
X$ is
$$
p(y)=\frac{\Gamma(\alpha+\beta)}{\Gamma(\alpha)\Gamma(\beta)}
\frac{\exp(\alpha y)}{(1+\exp(y))^{\alpha+\beta}},
$$
with moment generating function
$$
M_Y(z)=\E(\exp(zY))=
\frac{\Gamma(\alpha+\beta)}{\Gamma(\alpha)\Gamma(\beta)}
\int_{-\infty}^{\infty}
\frac{\exp((\alpha+z)y)}{(1+\exp(y))^{\alpha+\beta}}\,dy=
\frac{\Gamma(\alpha+z)\Gamma(\beta-z)}{\Gamma(\alpha)\Gamma(\beta)},
$$
for $z<\beta$. The cumulant generating function is $K(z)=\log
M_Y(z)=\log\Gamma(\alpha+z)+\log\Gamma(\beta-z)-\log\Gamma(\alpha)-\log\Gamma(\beta)$
and so
\begin{equation}\label{log:beta1}
\E(Y)=\left. \frac{\,dK(z)}{\,dz} \right|_{z=0} =
\psi(\alpha)-\psi(\beta) \quad \textrm{and} \quad \var(Y)=\left.
\frac{\,d^2K(z)}{\,dz^2} \right|_{z=0} =
\frac{\,d\psi(\alpha)}{\,d\alpha} + \frac{\,d\psi(\beta)}{\,d\beta}.
\end{equation}

For computational purposes, for large $x$, we can approximate
$\psi(x)$ by $\log x$ and $\,d\psi(x)/\,dx$ by $1/x$ (Abramowitz and
Stegun, 1964).

Thus, for the calculation of $r_t$ and $s_t$ in equation
(\ref{eq:binom:rt}), from the prior $\pi_t|y^{t-1}\sim
B(r_t,s_t-r_t)$, we have
\begin{equation}\label{rt:solv1}
f_t=\E(\eta_t|y^{t-1})=\E\left(\log
\frac{\pi_t}{1-\pi_t}\Big|y^{t-1}\right)=\psi(r_t)-\psi(r_t-s_t)=\log\frac{r_t}{s_t-r_t}
\end{equation}
and
\begin{equation}\label{st:solv1}
q_t=\var(\eta_t|y^{t-1})=\var\left(\log
\frac{\pi_t}{1-\pi_t}\Big|y^{t-1}\right)=\frac{\,d\psi(r_t)}{\,dr_t}-
\frac{\,d\psi(s_t-r_t)}{\,d(s_t-r_t)}=\frac{1}{r_t}-\frac{1}{s_t-r_t}
\end{equation}
We obtain (\ref{eq:binom:rt}) by solving (\ref{rt:solv1}) and
(\ref{st:solv1}) for $r_t$ and $s_t$.

The calculation of $r_t$ and $s_t$ of (\ref{eq:poisson:rt}) follows
a similar pattern. To this end, we note the gamma prior
$\lambda_t\sim G(r_t,s_t)$ and with the logarithmic link we have
\begin{equation}\label{rt:solv2}
f_t=\E(\eta_t|y^{t-1})=\E(\log\lambda_t|y^{t-1})=\psi(r_t)-\log(s_t)=\log
\frac{r_t}{s_t}
\end{equation}
and
\begin{equation}\label{st:solv2}
q_t=\var(\eta_t|y^{t-1})=\var(\log\lambda_t|y^{t-1})=\frac{\,d\psi(r_t)}{\,dr_t}=\frac{1}{r_t}.
\end{equation}
Equation (\ref{eq:poisson:rt}) is obtained by the solution of
(\ref{rt:solv2}) and (\ref{st:solv2}) for $r_t$ and $s_t$.

Since $f_t$ and $q_t$ are only guides of the mean and variance of
the prior of $\eta_t$, the above approximations of $\psi(x)$ and
$\,d\psi(x)/\,dx$ can be used even when $x$ is small. The posterior
quantities $f_t^*=\E(\eta_t|y^t)$ and $q_t^*=\var(\eta_t|y^t)$ are
calculated in a similar way, but here we use the full approximations
$\psi(x)=\log x+x^{-1}$ and $\,d\psi(x)/\,dx=x^{-1}(1-(2x)^{-1})$,
the details of which can be found in Abramowitz and Stegun (1964).

\subsection*{Proof of the prior (\ref{eq:igaussian:2}) and the expectation (\ref{IG:prior:m})}\label{sm2}

The prior distribution of $\gamma_t$ is
\begin{equation}\label{app1}
p(\gamma_t|y^{t-1})=\kappa(r_t,s_t)
\exp(r_t\gamma_t+2s_t\sqrt{-\gamma_t}).
\end{equation}
This is not a known distribution and so we need to use integration
in order to find the constant $\kappa(r_t,s_t)$. Since $\gamma_t<0$,
we need to evaluate
$$
I=\int_{-\infty}^0 \exp(r_t\gamma_t+2s_t\sqrt{-\gamma_t})
\,d\gamma_t
$$
By applying the substitution $y=\sqrt{-\gamma_t}$ we have
$$
I=2\exp\left(\frac{s_t^2}{r_t}\right) \int_0^{\infty} \exp \left(
-r_t \left( y - \frac{s_t}{r_t} \right)^2 \right) y \,d y =
2\exp\left(\frac{s_t^2}{r_t}\right) I_1.
$$
Now $I_1$ can be written as
$$
I_1=\frac{s_t}{r_t}\int_0^{\infty} \exp \left( -r_t \left( y -
\frac{s_t}{r_t} \right)^2 \right)\,dy + \int_0^{\infty}\exp \left(
-r_t \left( y - \frac{s_t}{r_t} \right)^2 \right) \left(
y-\frac{s_t}{r_t}\right) \,dy = I_2+I_3.
$$
Integral $I_2$ can be evaluated via the Gaussian integral, i.e.
$$
I_2=\frac{s_t}{2r_t} \int_{-\infty}^{\infty} \exp \left( -
\frac{\left( y -
\frac{s_t}{r_t}\right)^2}{\frac{2}{2r_t}}\right)\,dy=\frac{s_t}{2r_t}
\sqrt{ \frac{\pi}{r_t}}.
$$
For $I_3$ we use the substitution $(y-s_t/r_t)^2=z$ and so we get
$$
I_3=\frac{1}{2}\int_{s_t^2/r_t^2}^{\infty} \exp(-r_tz)\,dz =
\frac{1}{2r_t} \exp\left(-\frac{s_t^2}{r_t}\right).
$$
Thus, combining $I_1$, $I_2$ and $I_3$, we obtain
$$
\kappa(r_t,s_t)=I^{-1}=r_t\left( \exp\left(\frac{s_t^2}{r_t}\right)
s_t \sqrt{\frac{\pi}{r_t}} + 1\right)^{-1}.
$$
The required prior distribution of $\mu_t$ is immediately obtained
by density (\ref{app1}), if we apply the transformation
$\gamma_t=-1/\mu_t^2$ and we use $\kappa(r_t,s_t)$ as above.

Proceeding with the proof of (\ref{IG:prior:m}) we have
$$
\E(\mu_t|y^{t-1})=\int_{-\infty}^\infty \mu_t
p(\mu_t|y^{t-1})\,d\mu_t = c \int_0^\infty \frac{1}{\mu_t^2}
\exp\left(-\frac{(r_t-\mu s_t)^2}{r_t\mu_t^2}\right) = cI,
$$
where $c=(2\exp(s_t^2/r_t) r_t)/ (\exp(s_t^2/r_t) s_t \sqrt{\pi/r_t}
+ 1)$. To evaluate integral $I$ we note that
$(r_t-\mu_ts_t)^2/(r_t\mu_t^2)=r_t^{-1}(r_t\mu_t^{-1}-s_t)^2$ and by
applying the substitution $\mu_t^{-1}=-y$ and using the Gaussian
integral, we have
$$
I = \int_{-\infty}^0 \exp\left(-\frac{1}{r_t}
(r_ty+s_t)^2\right)\,dy  = \int_{-\infty}^0
\exp\left(-\frac{(y+s_tr_t^{-1})^2}{1/r_t}\right)\,dy  =
\frac{1}{2}\sqrt{\frac{\pi}{r_t}}.
$$
The required mean (\ref{IG:prior:m}) is obtained as $cI$.


\begin{thebibliography}{10}

\bibitem{Abramowitz}
Abramowitz, M. and Stegun, I.A. (1964) {\it Handbook of Mathematical
Functions.} Dover Publications, New York.

\bibitem{AitBr}
Aitchison, J. and Brown, J.A.C. (1957) {\it The Lognormal
Distribution: With Special Reference to Its Uses in Economics.}
Cambridge University Press, New-York.

\bibitem{Arnold}
Arnold, B.C. and Press, S.J. (1989) Bayesian estimation and
prediction for Pareto data. {\it Journal of the American Statistical
Association}, {\bf 84}, 1079-1084.

\bibitem{Chhikara}
Chhikara, R. and Folks, L. (1989) {\it The Inverse Gaussian
Distribution: Theory, Methodology, and Applications.} Marcel Dekker,
New-York.

\bibitem{Chiogna}
Chiogna, M. and Gaetan, C. (2002) Dynamic generalized linear models
with application to enironmental epidemiology. {\it Applied
Statistics}, {\bf 51}, 453-468.

\bibitem{Chong04}
Chong, J. (2004) Value at Risk from econometric models and implied
from currency options. {\it Journal of Forecasting}, {\bf 23},
603-620.

\bibitem{Cox}
Cox, D.R. (1981) Statistical analysis of time-series: some recent
developments. {\it Scandanavian Journal of Statistics}, {\bf 8},
93-115.

\bibitem{Dobson}
Dobson, A.J. (2002) {\it An Introduction to Generalized Linear
Models.} 2nd edition, Chapman and Hall, New York.

\bibitem{Durbin00}
Durbin, J. and Koopman, S.J. (2000) Time series analysis of
non-Gaussian observations based on state space models from both
classical and Bayesian perspectives (with discussion). {\it Journal
of the Royal Statistical Society Series B}, {\bf{62}}, 3-56.

\bibitem{Durbin01}
Durbin, J. and Koopman, S.J. (2001) {\it Time Series Analysis by
State Space Methods}. Oxford University Press, Oxford.

\bibitem{Dyer}
Dyer, D. (1981) Structural probability bounds for the strong Pareto
law. {\it Canadian Journal of Statistics}, {\bf 9}, 71-77.

\bibitem{Fahrmeir87}
Fahrmeir, L. (1987) Regression models for nonstationary categorical
time series. {\it Journal of Time Series Analysis}, {\bf 8},
147-160.

\bibitem{Fahrmeir94}
Fahrmeir, L. and Tutz, G. (2001) {\it Multivariate Statistical
Modelling Based on Generalized Linear Models}. 2nd edition,
Springer, New York.

\bibitem{Ferreira}
Ferreira, M.A.R. and Gamerman, D. (2000) Dynamic generalized linear
models. In {\it Generalized Linear Models: A Bayesian Perspective},
D.K. Dey, S.K. Ghosh and B.K. Mallick (Eds.). Marcel Dekker, new
York.

\bibitem{Schnatter94}
Fr\"{u}hwirth-Schnatter, S. (1994) Applied state space modelling of
non-Gaussian time series using integration-based Kalman filtering.
{\it Statistics and Computing}, {\bf 4}, 259-269.

\bibitem{Gamerman91}
Gamrman, D. (1991) Dynamic Bayesian models for survival data. {\it
Applied Statistics}, {\bf 40}, 63-79.

\bibitem{Gamerman98}
Gamrman, D. (1998) Markov chain Monte Carlo for dynamic generalised
linear models. {\it Biometrika}, {\bf 85}, 215-227.

\bibitem{GamermanWest}
Gamerman, D. and West, M. (1987) An application of dynamic survival
models in unemployment studies. {\it Statistician}, {\bf 36},
269-274.

\bibitem{Godolphin06}
Godolphin, E.J. and Triantafyllopoulos, K. (2006) Decomposition of
time series models in state-space form. {\it Computational
Statistics and Data Analysis}, {\bf 50}, 2232-2246.

\bibitem{Harvey04}
Harvey, A.C. (2004) Tests for cycles. In {\it State Space and
Unobserved Component Models: Theory and Applications}, A.C Harvey,
S.J. Koopman and N. Shephard (Eds.). Cambridge University Press,
Cambridge.

\bibitem{Hemming}
Hemming, K. and Shaw, J.E.H. (2002) A parametric dynamic survival
model applied to breast cancer survival times. {\it Applied
Statistics}, {\bf 51}, 421-435.

\bibitem{Housman}
Houseman, E.A., Coull, B.A. and Shine, J.P. (2006) A nonstationary
negative binomial time series with time-dependent covariates:
enterococcus counts in Boston harbor. {\it Journal of the American
Statistical Association}, {\bf 101}, 1365-1376.

\bibitem{Huberman}
Huberman, B.A., Pirolli, P.L.T., Pitkow, J.E. and Lukose, R.M.
(1998) Strong regularities in world wide web surfng. {\it Science},
{\bf 280}, 95-97.

\bibitem{Johnson05}
Johnson, N.L., Kemp, A.W and Kotz, S. (2005) {\it Univariate
Discrete Distributions.} 3rd edition, Wiley, New-York.

\bibitem{Johnson94}
Johnson, N.L., Kotz, S. and Balakrishnan, N. (1994) {\it Continuous
Univariate Distributions, Volume 1.} 2nd edition, Wiley, New-York.

\bibitem{Jung}
Jung, R.C., Kukuk, M. and Liesenfeld, R. (2006) Time series of count
data: modeling, estimation and diagnostics. {\it      Computational
Statistics and Data Analysis}, {\bf 51}, 2350-2364.

\bibitem{Kaufmann}
Kaufmann, H. (1987) Regression models for nonstationary categorical
time series: asymptotic estimation theory. {\it Annals of
Statistics}, {\bf 15}, 79-98.

\bibitem{Kedem}
Kedem, B. and Fokianos, K. (2002) {\it Regression Models for Time
Series Analysis.} Wiley, new York.

\bibitem{Kendal90}
Kendall, M.G. and Ord, J.K. (1990) {\it Time Series}. 3rd edition,
Edward Arnold.

\bibitem{Kitagawa87}
Kitagawa, G. (1987) Non-Gaussian state-space modelling of
nonstationary time series. {\it Journal of the American Statistical
Association}, {\bf 82}, 1032-1063.

\bibitem{Limpert}
Limpert, E., Stahel, W.A. and Abbt, M. (2001) Lognormal
distributions across the sciences: keys and clues. {\it Bioscience}
{\bf 51}, 341-352.

\bibitem{Lindsey}
Lindsey, J.K. and Lambert, P. (1995) Dynamic generalized linear
models and repeated measurements. {\it Journal of Statistical
Planning and Inference}, {\bf 47}, 129-139.

\bibitem{Lindsey97}
Lindsey, J.K. (1997) {\it Applying Generalized Linear Models.}
Springer, New York.

\bibitem{McCullagh89}
McCullagh, P. and Nelder, J.A. (1989) {\it Generalized Linear
Models}. 2nd edition, Chapman and Hall, London.

\bibitem{Morrison}
Morrison, J. (1958) The lognormal distribution in quality control.
{\it Applied Statistics}, {\bf 7}, 160-172.

\bibitem{nandram}
Nandram, B. and Kim, H. (2002) Marginal likelihood for a class of
Bayesian generalized linear models. {\it Journal of Statistical
Computation and Simulation}, {\bf 72}, 319-340.

\bibitem{Salvador2}
Salvador, M. and Gargallo, P. (2005) Automatic selective
intervention in dynamic linear models . {\it Journal of Applied
Statistics}, {\bf 30}, 1161-1184.

\bibitem{Shephard97}
Shephard, N. and Pitt, M.K. (1997) Likelihood analysis of
non-Gaussian measurement time series. {\it Biometrika}, {\bf 84},
653-667.

\bibitem{Smith79}
Smith, J.Q. (1979) A generalization of the Bayesian steady
forecasting model. {\it Journal of the Royal Statistical Society
Series B}, {\bf 41}, 375-387.

\bibitem{Takagi}
Takagi, K., Kumaga, S., Matsunaga, I. and Kusaka, Y. (1997)
Application of inverse Gaussian distribution to occupational
exposure data. {\it Annals of Occupational Hygiene}, {\bf 41},
505-514.

\bibitem{Tsay02}
Tsay, R.S. (2002). {\it Analysis of Financial Time Series}. Wiley,
New York.

\bibitem{West97}
West, M. and Harrison, P.J. (1997) {\it Bayesian Forecasting and
Dynamic Models}. 2nd edition, Springer, New York.

\bibitem{West85}
West, M., Harrison, P.J. and Migon, H.S. (1985) Dynamic generalized
linear models and Bayesian forecasting (with discussion). {\it
Journal of the American Statistical Association}, {\bf 80}, 73-97.

\end{thebibliography}
\end{document}